\documentclass[sigconf]{acmart}
\AtBeginDocument{%
  \providecommand\BibTeX{{%
    \normalfont B\kern-0.5em{\scshape i\kern-0.25em b}\kern-0.8em\TeX}}}

\usepackage[acronyms,nonumberlist,nopostdot,nomain,nogroupskip]{glossaries}
\usepackage{tablefootnote}
\newacronym{ca}{CA}{Carrier Aggregation}
\newacronym{3gpp}{3GPP}{3rd Generation Partnership Project}
\newacronym{5g}{5G}{5th generation}
\newacronym{5gc}{5GC}{5G Core}
\newacronym{adc}{ADC}{Analog to Digital Converter}
\newacronym{afbw}{AFBW}{Average Fading Bandwidth}
\newacronym{aimd}{AIMD}{Additive Increase Multiplicative Decrease}
\newacronym{am}{AM}{Acknowledged Mode}
\newacronym{amc}{AMC}{Adaptive Modulation and Coding}
\newacronym{aoa}{AoA}{Angle of Arrival}
\newacronym{aod}{AoD}{Angle of Departure}
\newacronym{ap}{AP}{Access Point}
\newacronym{af}{AF}{Amplify-and-Forward}
\newacronym{rr}{RR}{regenerative relay}
\newacronym{aqm}{AQM}{Active Queue Management}
\newacronym{awgn}{AGWN}{Additive White Gaussian Noise}
\newacronym{balia}{BALIA}{Balanced Link Adaptation}
\newacronym{bdp}{BDP}{Bandwidth-Delay Product}
\newacronym{ber}{BER}{Bit Error Rate}
\newacronym{bf}{BF}{Beamforming}
\newacronym{bwp}{BWP}{Bandwidth Part}
\newacronym{cad}{CAD}{Computer-Aided Design}
\newacronym{cc}{CC}{Congestion Control}
\newacronym{cdf}{CDF}{Cumulative Distribution Function}
\newacronym{cir}{CIR}{Channel Impulse Response}
\newacronym{cn}{CN}{Core Network}
\newacronym{cp}{CP}{Control Plane}
\newacronym{cqi}{CQI}{Channel Quality Information}
\newacronym{crs}{CRS}{Cell Reference Signal}
\newacronym{csirs}{CSI-RS}{Channel State Information - Reference Signal}
\newacronym{dc}{DC}{Dual Connectivity}
\newacronym{dce}{DCE}{Direct Code Execution}
\newacronym{dci}{DCI}{Downlink Control Information}
\newacronym{dl}{DL}{Downlink}
\newacronym{dmr}{DMR}{Deadline Miss Ratio}
\newacronym{dmrs}{DMRS}{DeModulation Reference Signal}
\newacronym{dray}{D-Ray}{Deterministic Ray}
\newacronym{e2e}{E2E}{end-to-end}
\newacronym{ecn}{ECN}{Explicit Congestion Notification}
\newacronym{edf}{EDF}{Earliest Deadline First}
\newacronym{em}{EM}{electromagnetic}
\newacronym{enb}{eNB}{evolved Node Base}
\newacronym{endc}{EN-DC}{E-UTRAN-\gls{nr} \gls{dc}}
\newacronym{epc}{EPC}{Evolved Packet Core}
\newacronym{es}{ES}{Edge Server}
\newacronym{fdd}{FDD}{Frequency Division Duplexing}
\newacronym{fdma}{FDMA}{Frequency Division Multiple Access}
\newacronym{fray}{F-Ray}{Flashing Ray}
\newacronym{fs}{FS}{Fast Switching}
\newacronym{ftp}{FTP}{File Transfer Protocol}
\newacronym{gmm}{GMM}{Gaussian Mixture Model}
\newacronym{gnb}{gNB}{Next Generation Node Base}
\newacronym{harq}{HARQ}{Hybrid Automatic Repeat reQuest}
\newacronym{hetnet}{HetNet}{Heterogeneous Network}
\newacronym{hh}{HH}{Hard Handover}
\newacronym{hol}{HOL}{Head-of-Line}
\newacronym{hqf}{HQF}{Highest-quality-first}
\newacronym{ia}{IA}{Initial Access}
\newacronym{iab}{IAB}{Integrated Access and Backhaul}
\newacronym{imt}{IMT}{International Mobile Telecommunication}
\newacronym{inf}{InF}{Indoor Factory}
\newacronym{irs}{IRS}{Intelligent Reflective Surface}
\newacronym{inr}{INR}{Interference to Noise Ratio}
\newacronym{iot}{IoT}{Internet of Things}
\newacronym{ked}{KED}{Knife-Edge Diffraction}
\newacronym{kpi}{KPI}{Key Performance Indicator}
\newacronym{lcf}{LCF}{Level Crossing Frequency}
\newacronym{lcr}{LCR}{Level Crossing Rate}
\newacronym{los}{LoS}{Line-of-Sight}
\newacronym{l2sm}{L2SM}{Link-to-System Mapping}
\newacronym{lte}{LTE}{Long Term Evolution}
\newacronym{ltemtp}{LTE-M}{LTE-MTC [Machine Type Communication]}
\newacronym{m2m}{M2M}{Machine to Machine}
\newacronym{mac}{MAC}{Medium Access Control}
\newacronym{mc}{MC}{Monte Carlo}
\newacronym{mcs}{MCS}{Modulation and Coding Scheme}
\newacronym{mec}{MEC}{Mobile Edge Cloud}
\newacronym{mi}{MI}{Mutual Information}
\newacronym{mib}{MIB}{Master Information Block}
\newacronym{mimo}{MIMO}{Multiple Input Multiple Output}
\newacronym{siso}{SISO}{Single Input Single Output}
\newacronym{m-mimo}{m-MIMO}{massive MIMO}
\newacronym{mlr}{MLR}{Maximum-local-rate}
\newacronym{ftr}{FTR}{Fluctuating Two-Ray}
\newacronym{mmwave}{mmWave}{millimeter wave}
\newacronym{moi}{MoI}{Method of Images}
\newacronym{pdf}{PDF}{Probability Density Function}
\newacronym{mpc}{MPC}{Multi Path Component}
\newacronym{mptcp}{MPTCP}{Multipath TCP}
\newacronym{mr}{MR}{Maximum Rate}
\newacronym{mrdc}{MR-DC}{Multi \gls{rat} \gls{dc}}
\newacronym{mss}{MSS}{Maximum Segment Size}
\newacronym{mtd}{MTD}{Machine-Type Device}
\newacronym{mtu}{MTU}{Maximum Transmission Unit}
\newacronym{nfv}{NFV}{Network Function Virtualization}
\newacronym{nist}{NIST}{National Institute of Standards and Technology}
\newacronym{nlos}{NLoS}{Non-Line-of-Sight}
\newacronym{ntn}{NTN}{Non-Terrestrial Networks}
\newacronym{nr}{NR}{New Radio}
\newacronym{nrmse}{NRMSE}{Normalized Root Mean Square Error}
\newacronym{nsa}{NSA}{Non Stand Alone}
\newacronym{o2i}{O2I}{Outdoor-to-Indoor}
\newacronym{ofdm}{OFDM}{Orthogonal Frequency Division Multiplexing}
\newacronym{pa}{PA}{power amplifier}
\newacronym{prr}{PRR}{Packet Reception Ratio}
\newacronym{pbch}{PBCH}{Physical Broadcast Channel}
\newacronym{pdcch}{PDCCH}{Physical Downlonk Control Channel}
\newacronym{pdcp}{PDCP}{Packet Data Convergence Protocol}
\newacronym{pdsch}{PDSCH}{Physical Downlink Shared Channel}
\newacronym{pdu}{PDU}{Packet Data Unit}
\newacronym{per}{PER}{Packet Error Rate}
\newacronym{pf}{PF}{Proportional Fair}
\newacronym{pgw}{PGW}{Packet Gateway}
\newacronym{phy}{PHY}{Physical}
\newacronym{pl}{PL}{Path Loss}
\newacronym{ppp}{PPP}{Poisson Point Process}
\newacronym{prb}{PRB}{Physical Resource Block}
\newacronym{psd}{PSD}{Power Spectral Density}
\newacronym{pss}{PSS}{Primary Synchronization Signal}
\newacronym{pucch}{PUCCH}{Physical Uplink Control Channel}
\newacronym{pusch}{PUSCH}{Physical Uplink Shared Channel}
\newacronym{qam}{QAM}{Quadrature Amplitude Modulation}
\newacronym{qd}{QD}{Quasi Deterministic}
\newacronym{rach}{RACH}{Random Access Channel}
\newacronym{ran}{RAN}{Radio Access Network}
\newacronym[firstplural=Radio Access Technologies (RATs)]{rat}{RAT}{Radio Access Technology}
\newacronym{red}{RED}{Random Early Detection}
\newacronym{rf}{RF}{Radio Frequency}
\newacronym{rlc}{RLC}{Radio Link Control}
\newacronym{rlf}{RLF}{Radio Link Failure}
\newacronym{rray}{R-Ray}{Random Ray}
\newacronym{rrc}{RRC}{Radio Resource Control}
\newacronym{rrm}{RRM}{Radio Resource Management}
\newacronym{rs}{RS}{Remote Server}
\newacronym{rsrp}{RSRP}{Reference Signal Received Power}
\newacronym{rsrq}{RSRQ}{Reference Signal Received Quality}
\newacronym{rss}{RSS}{Received Signal Strength}
\newacronym{rssi}{RSSI}{Received Signal Strength Indicator}
\newacronym{rt}{RT}{Ray Tracer}
\newacronym{rtt}{RTT}{Round Trip Time}
\newacronym{rw}{RW}{Receive Window}
\newacronym{rx}{RX}{Receiver}
\newacronym{sa}{SA}{standalone}
\newacronym{sack}{SACK}{Selective Acknowledgment}
\newacronym{sap}{SAP}{Service Access Point}
\newacronym{sch}{SCH}{Secondary Cell Handover}
\newacronym{scm}{SCM}{Spatial Channel Model}
\newacronym{scoot}{SCOOT}{Split Cycle Offset Optimization Technique}
\newacronym{sdap}{SDAP}{Service Data Adaptation Protocol}
\newacronym{sdma}{SDMA}{Spatial Division Multiple Access}
\newacronym{sf}{SF}{Shadow Fading}
\newacronym{si}{SI}{Study Item}
\newacronym{sib}{SIB}{Secondary Information Block}
\newacronym{sinr}{SINR}{Signal-to-Interfe\-rence-plus-Noise Ratio}
\newacronym{sir}{SIR}{Signal-to-Interference Ratio}
\newacronym{sm}{SM}{Saturation Mode}
\newacronym{snr}{SNR}{Signal-to-Noise Ratio}
\newacronym{son}{SON}{Self-Organizing Network}
\newacronym{srs}{SRS}{Sounding Reference Signal}
\newacronym{ss}{SS}{Synchronization Signal}
\newacronym{sss}{SSS}{Secondary Synchronization Signal}
\newacronym{sta}{STA}{Station}
\newacronym{tb}{TB}{Transport Block}
\newacronym{tcp}{TCP}{Transmission Control Protocol}
\newacronym{udp}{UDP}{User Datagram Protocol}
\newacronym{tdd}{TDD}{Time Division Duplexing}
\newacronym{tdma}{TDMA}{Time Division Multiple Access}
\newacronym{tfl}{TfL}{Transport for London}
\newacronym{tgad}{TGad}{Task Group ad}
\newacronym{tgay}{TGay}{Task Group ay}
\newacronym{tm}{TM}{Transparent Mode}
\newacronym{thz}{THz}{Terahertz}
\newacronym{trp}{TRP}{Transmitter Receiver Pair}
\newacronym{tti}{TTI}{Transmission Time Interval}
\newacronym{ttt}{TTT}{Time-to-Trigger}
\newacronym{tx}{TX}{Transmitter}
\newacronym{ue}{UE}{User Equipment}
\newacronym{ul}{UL}{Uplink}
\newacronym{um}{UM}{Unacknowledged Mode}
\newacronym{uma}{UMa}{Urban Macro}
\newacronym{uml}{UML}{Unified Modeling Language}
\newacronym{utc}{UTC}{Urban Traffic Control}
\newacronym{vm}{VM}{Virtual Machine}
\newacronym{wbf}{WBF}{Wired Bias Function}
\newacronym{wf}{WF}{Wired-first}
\newacronym{wifi}{Wi-Fi}{Wireless Fidelity}
\newacronym{wigig}{WiGig}{Wireless Gigabit}
\newacronym{wlan}{WLAN}{Wireless Local Area Network}
\newacronym{xpr}{XPR}{Cross Polarization Ratio}
\newacronym{fr2}{FR2}{Frequency Range 2}
\newacronym{nbiot}{NB-IoT}{Narrowband-IoT}
\newacronym{cps}{CPS}{Cyber-Physical production System}
\newacronym{iiot}{IIoT}{Industrial Internet of Things}
\newacronym{agv}{AGV}{Autonomous Ground Vehicle}
\newacronym{uav}{UAV}{Unmanned Autonomous Vehicle}
\newacronym{amr}{AMR}{Autonomous Mobile Robots}
\newacronym{wsn}{WSN}{Wireless Sensor Network}
\newacronym{embb}{eMBB}{enhanced Mobile Broadband}
\newacronym{urllc}{URLLC}{Ultra-Reliable Low-Latency Communications}
\newacronym{xr}{XR}{Extended Reality}
\newacronym{ai}{AI}{Artificial Intelligence}
\newacronym{ml}{ML}{Machine Learning}
\newacronym{drl}{DRL}{Deep Reinforcement Learning}
\newacronym{uas}{UAS}{Unmanned Aircraft System}
\newacronym{stl}{STL}{Standard Template Library}
\newacronym{svd}{SVD}{Singular Value Decomposition}
\newacronym{itu}{ITU}{International Telecommunications Union}
\newacronym{upa}{UPA}{Uniform Planar Array}
\newacronym{simd}{SIMD}{Single instruction, multiple data}
\newacronym{eesm}{EESM}{Exponential Effective SNR Mapping}
\newacronym{l2s}{L2S}{Link-to-System}

\usepackage{subcaption}
\usepackage{tikz}
\usepackage{pgfplots}
\pgfplotsset{compat=newest}
\pgfplotsset{plot coordinates/math parser=false}
\newlength\fheight
\newlength\fwidth
\usetikzlibrary{plotmarks,decorations,backgrounds,calc,spy}
\usetikzlibrary{decorations.markings}
\usepgfplotslibrary{patchplots,groupplots}
\usepackage{tikzscale}
\usetikzlibrary{spy}

\DeclareMathOperator*{\argmin}{argmin}

\copyrightyear{2023}
\acmYear{2023}
\setcopyright{acmlicensed}\acmConference[WNS3 2023]{2023 Workshop on ns-3}{June 28--29, 2023}{Arlington, VA, USA}
\acmBooktitle{2023 Workshop on ns-3 (WNS3 2023), June 28--29, 2023, Arlington, VA, USA}
\acmPrice{15.00}
\acmDOI{10.1145/3592149.3592167}
\acmISBN{979-8-4007-0747-6/23/06}

\usepackage{bm}
\usepackage{soul}

\hypersetup{draft}

\begin{document}

\title{Improving the Efficiency of MIMO Simulations in ns-3}

\flushbottom
\setlength{\parskip}{0ex plus0.1ex}
\addtolength{\skip\footins}{-0.2pc plus 40pt}

\author{Matteo Pagin}
\affiliation{%
  \institution{University of Padova}
  \city{Padova}
  \country{Italy}
}
\email{paginmatte@dei.unipd.it}

\author{Sandra Lagén}
\author{Biljana Bojović}
\affiliation{%
  \institution{Centre Tecnològic de Telecomunicacions de Catalunya}
  \city{Barcelona}
  \country{Spain}}
\email{{sandra.lagen, biljana.bojovic}@cttc.es}

\author{Michele Polese}
\affiliation{%
\institution{Institute for the Wireless Internet of Things,\\Northeastern University}
 \city{Boston}
 \country{MA, USA}}
\email{m.polese@northeastern.edu}

\author{Michele Zorzi}
\affiliation{%
  \institution{University of Padova}
  \city{Padova}
  \country{Italy}
}
\email{zorzi@dei.unipd.it}

\begin{abstract}
Channel modeling is a fundamental task for the design and evaluation of wireless technologies and networks, before actual prototyping, commercial product development and real deployments. 
The recent trends of current and future mobile networks, which include large antenna systems, massive deployments, and high-frequency bands, require complex channel models for the accurate simulation of \gls{m-mimo} in \gls{mmwave} and \gls{thz} bands. To address the complexity/accuracy trade-off, a spatial channel model has been defined by 3GPP (TR 38.901), which has been shown to be the main bottleneck of current system-level simulations in ns-3. 
In this paper, we focus on improving the channel modeling efficiency for large-scale MIMO system-level simulations. Extensions are developed in two directions. First, we improve the efficiency of the current 3GPP TR 38.901 implementation code in ns-3, by allowing the use of the Eigen library for more efficient matrix algebra operations, among other optimizations and a more modular code structure. Second, we propose a new performance-oriented MIMO channel model for reduced complexity, as an alternative model suitable for \gls{mmwave}/\gls{thz} bands, and calibrate it against the 3GPP TR 38.901 model. Simulation results demonstrate the proper calibration of the newly introduced model for various scenarios and channel conditions, and exhibit an effective reduction of the simulation time (up to 16 times compared to the previous baseline) thanks to the various proposed improvements. 
\end{abstract}




\maketitle

\section{Introduction}

Mobile networks play a key role in our society and are poised to become ever more important in the coming years. In fact, the \gls{itu} foresees that in 2030 and beyond wireless broadband will be ubiquitous, and will be required to provide connectivity not only to humans, but also to a plethora of intelligent devices such as wearables, road vehicles, \glspl{uas} and robots~\cite{imt2030}. Moreover, novel use cases such as holographic communications, \gls{xr} and tactile applications will further exacerbate the throughput and latency requirements which were posed by \gls{embb} and \gls{urllc}~\cite{itu-r-2083}. 


To meet these goals, future cellular systems will 
further evolve \gls{5g} 
networks,
which have introduced a flexible, virtualized architecture, the support for \gls{mmwave} communications and the use of \gls{m-mimo} technologies~\cite{ghosh20195g}. Notably, the research community is considering a more central role for \glspl{mmwave}, a further expansion of the spectrum towards the \gls{thz} band, and an \gls{ai}-native network design, with the goal of achieving autonomous data-centric orchestration and management of the network~\cite{polese20216g}, possibly down to the air interface~\cite{hoydis2021toward}.

The \gls{thz} and \gls{mmwave} bands offer large chunks of untapped bandwidth which operators can leverage to meet the Tb/s peak rates that are envisioned by the ITU~\cite{imt2030}. However, this portion of the spectrum is plagued by unfavorable propagation characteristics, comprising a marked free-space propagation loss 
and susceptibility to blockages~\cite{han2018propagation, jornet2011channel}, which make it challenging to harvest its potential. Although the harsh propagation environment can be partially mitigated by using directional links 
and densifying network deployments~\cite{polese2020toward}, 
the support for \gls{mmwave} and \gls{thz} bands entails a major redesign not only of the physical layer, but of the whole cellular protocol stack~\cite{shafi2018microwave}. For instance, the intrinsic directionality of the communication requires ad hoc control procedures~\cite{heng2021six},
while the frequent transitions between \gls{los} and \gls{nlos} conditions call for an ad hoc transport layer design, such as novel \gls{tcp} algorithms~\cite{zhang2019will}. 

In addition, as the network progressively becomes increasingly complex and heterogeneous, the push for spectrum expansion will be coupled with an \gls{ai}-native design which, thanks to the ongoing virtualization, will not be limited to the radio link level, but will encompass the orchestration of large scale deployments as well~\cite{polese2023understanding}.
Nevertheless, how to design, test and eventually deploy management and orchestration algorithms is an open research challenge~\cite{polese2022colo}.
First, the training data must accurately capture the interplay of the whole protocol stack with the wireless channel. Furthermore, optimization frameworks such as \gls{drl} also call for preliminary testing in isolated yet realistic environments, with the goal of minimizing the performance degradation to actual network deployments~\cite{lacava2022programmable, amir2023safehaul}.

In these regards, system-level network simulators have a central role to play. Indeed, an end-to-end evaluation of algorithms and protocols becomes paramount when considering frequencies above 6~GHz, given the impact of their peculiar propagation characteristics on the whole protocol stack.
At the same time, end-to-end simulators can also serve as both sources of training data for \gls{ai} models, and testing platforms for preliminary evaluation of \gls{ml} algorithms prior to their deployment in commercial networks.
However, the suitability of end-to-end network simulators to these tasks largely depends on the accuracy of the channel model~\cite{testolina2020scalable} and on the scalability for realistically-sized deployments.
In fact, system-level simulators generally abstract the actual link-level transmission via an error model, which maps the \gls{sinr} of the wireless link to a packet error probability~\cite{lagen2020new}. Eventually, the latter is used to determine whether the packet has been successfully decoded by the receiver. As a consequence, the accuracy of the simulator heavily depends on the reliability of the \gls{sinr} estimation, especially when considering the \gls{mmwave} and \gls{thz} bands. 

The well known ns-3 simulator features the implementation of the \gls{3gpp} channel model~\cite{TR38901}, which, according to the \gls{3gpp}, represents the state-of-the-art channel model for drop-based end-to-end simulations of devices operating at frequencies between 0.5 to 100~GHz. Despite its accuracy, the TR 38.901 channel model is particularly demanding from a computational point of view, and thus limits the scalability of the simulated scenarios. 
At the same time, the simpler channel models which are found in analytical studies fail to capture the peculiar characteristics of \gls{mmwave} and \gls{thz} links. 

To fill this gap, in this paper we propose optimizations to the ns-3 implementation of the TR 38.901 channel model of~\cite{tommaso:20}, both at the codebase and at the design level, which aim to provide wireless researchers with the tools for simulating future dense wireless scenarios in a computationally efficient manner. Specifically, we significantly improve the runtime of simulations involving the 3GPP TR 38.901 channel model~\cite{TR38901} by porting the intensive linear algebra operations to the open-source library \texttt{Eigen}~\cite{eigenweb}. To this end, we also design and implement a set of common linear algebra APIs, which increase the modularity of the \texttt{spectrum} module with respect to the underlying data structures and algorithms.
Then, we propose a simplified channel model, based on~\cite{TR38901}, which aims to provide an additional order of magnitude of runtime reduction, at the cost of a slight accuracy penalty. Profiling results show that the support for \texttt{Eigen}, coupled with further TR 38.901 optimizations, leads to a decrease of up to 5 times in the simulation time of typical \gls{mimo} scenarios. Furthermore, the proposed performance-oriented channel model further improved the runtime of simulations, which now take as low as $6$~\% with respect to the full TR 38.901 channel model, with a negligible loss in accuracy.

The remainder of the paper is organized as follows. Section~\ref{sec:rel_work} reports the state of the art on channel models. Sections~\ref{sec:opt_code} and~\ref{sec:opt_design} describe the contributions of this work, in terms of optimizations to the ns-3 implementation of the TR 38.901 framework and the design of a performance oriented channel model, respectively. Finally, Section~\ref{sec:results} presents benchmarks of the introduced optimizations and discusses the main use cases of these channel models, while Section~\ref{sec:conc} concludes the paper by mentioning possible future extensions of this work. 

\section{Related work}
\label{sec:rel_work}
Channel modeling is a fundamental task for the design, simulation, and evaluation of current and future wireless networks. It is especially relevant to perform system-level simulations to test new algorithms, procedures, and architectures, before going into real deployment/device implementations. In the recent decades, the challenges for understanding the propagation at \gls{mmwave} and \gls{thz} frequencies with large antenna arrays and the use of \gls{mimo} have further motivated the channel modeling efforts in those frequency ranges~\cite{hemadeh2018millimeter, 9444237}. As a result, multiple channel measurement campaigns have been performed by the academic and industry communities~\cite{rappaport2013millimeter}, 
leading to different families of channel models. The various channel models differ in their degree of simplicity and accuracy. 
They range from simple models that just consider a propagation loss component combined with Nakagami-m or Rayleigh fading but fail to capture the spatial dimension of the channel and the interactions with beamforming~\cite{andrews2017modeling}, to deterministic models that are very accurate in specific scenarios but are much more complex and require a precise characterization of the environment~\cite{lecci2020simplified}.
To address the complexity-accuracy trade-off, the \gls{3gpp} has adopted a stochastic channel model for simulations of 5G and beyond networks~\cite{TR38901}. Stochastic channel models are generic, thanks to their stochastic nature, but at the same time can model interactions with multiple-antenna arrays.
Specifically, the \gls{3gpp} defined in TR 38.901 the spatial channel model for simulations that address frequency ranges from 0.5 GHz to 100 GHz~\cite{TR38901}, which is parameterized for various simulation scenarios, including indoor office, indoor factory, urban macro, urban micro, and rural macro. 

However, for system-level simulations of large-scale systems including multiple nodes and large antenna arrays, the \gls{3gpp} spatial channel model still introduces a significant overhead in terms of computational complexity. In this line, in~\cite{8445856}, a simplified channel model for the system-level simulations of \gls{mimo} wireless networks is proposed. Therein, the end-to-end channel gain is obtained as the sum of several loss and gain terms that account for large-scale phenomena such as path loss and shadowing, small-scale fading, and antenna and beamforming gains. Notably, the latter terms represent a fundamental component for studies concerning modern wireless systems. In particular, an accurate characterization of the antenna radiation pattern and of the effect of the presence of multiple radiating elements becomes extremely important when studying \gls{mmwave} and \gls{thz} frequencies. Following the model in~\cite{8445856}, the combined antenna and beamforming gain can be computed according to~\cite{8422746}, the path loss and shadowing components can follow the \gls{3gpp} model in~\cite{TR38901}, and the small-scale fading can be sampled from various statistical distributions. 
For the small-scale fading, authors in~\cite{8445856} propose to use a Nakagami-$m$ distribution, which has been shown to provide a good fit with the 3GPP model, provided that the $m$ parameter is appropriately chosen.
Another option for small-scale fading modeling is the so-called \gls{ftr} fading model presented in~\cite{7917287}, which models more accurately the fading that occurs at mmWaves.

The 3GPP TR 38.901 spatial channel model was included in ns-3 thanks to the efforts of Tommaso Zugno in 2019 Google Summer of Code~\cite{tommaso:20}, and later extended to address vehicular scenarios in~\cite{10.1145/3460797.3460801} and industrial scenarios in~\cite{10.1145/3532577.3532596}. The current spatial channel model implemented in ns-3 is very accurate for simulations in line with \gls{3gpp} specifications for a wide range of frequencies, but represents the main bottleneck in terms of computational complexity when considering large-scale simulations with many multi-antenna nodes, especially when equipped with large antenna arrays. This is because of the intrinsic complexity in the generation of the channel model according to 3GPP specifications and the need to deal with 3D channel matrix structures. The channel matrix in the ns-3 implementation of the \gls{3gpp} spatial channel model is implemented as a 3D structure whose dimensions depend on the number of the transmit antennas, receive antennas, and clusters. Currently, in ns-3, the 3GPP channel model uses a vector of vectors of vectors to represent 3D arrays, such as the channel matrix. 

The design of computationally efficient yet accurate channel models has been a topic of interest also in the \gls{wlan} space. The authors of~\cite{jin2020efficient, jin2021efficient} present a frequency-selective channel for \glspl{wlan}, and use \gls{eesm} \gls{l2s} mapping to integrate their model with the ns-3 system-level Wi-Fi implementation. Moreover, they develop a framework which leverages cached statistical channel matrix realizations to directly estimate the effective \gls{snr}, thus further improving the computational efficiency of the model. Specifically, the latter is modeled as a parameterized log-SGN random variable. They extend their work in~\cite{jin2021eesm}, by accounting for the channel correlation over time. 
Moreover,~\cite{liu2021performance} compares statistical channel models for the 60~GHz band with the \gls{qd} \gls{rt} of~\cite{QD}.

In this paper, we summarize the efforts carried out by Matteo Pagin in 2022 Google Summer of Code to further optimize the code in ns-3 in two directions: 1) improving the efficiency of the code by allowing the use of \texttt{Eigen} library, and 2) proposing a new performance-oriented MIMO channel model for reduced complexity in ns-3 large-scale simulations. First, we have improved the efficiency of the \gls{3gpp} spatial channel model in ns-3 by allowing the usage of \texttt{Eigen} to represent matrices, so that when \texttt{Eigen} is available the \gls{3gpp} channel matrix is represented as an \texttt{std::vector} of \texttt{Eigen} matrices. This already improves the performance of current models. Second, we propose an alternative model, based on the FTR channel model~\cite{7917287}, in which the channel is characterized by a single scalar instead of 3D matrices, and we have calibrated such model to align with the \gls{3gpp} TR 38.901 spatial channel model for various scenarios and channel conditions. This model is especially useful to speed up ns-3 large-scale simulations, when simplicity is prioritized.

\section{Efficient MIMO modeling with the Eigen library}
\label{sec:opt_code}

The use of multiple antennas both at the transmitter and at the receiver, a fundamental feature of modern wireless systems, makes a scalar representation of the channel impulse response insufficient. Instead, \gls{mimo} channels are usually represented in the form of a complex matrix $\bm{H} \in \mathbb{C}^{U \times S}$, whose elements depict the channel impulse response between the $U$ and $S$ radiating elements of the transmitting and receiving antenna arrays, respectively~\cite{TR38901}. This peculiarity significantly increases the computational complexity of \gls{mimo} channel models, compared to \gls{siso} ones, since the complex gain of the channel must be evaluated for each pair of transmit and receive antennas.
Notably, previous analyses identified in statistical channel models the main bottleneck for system-level \gls{mimo} wireless simulations. In typical \gls{m-mimo} \gls{5g} scenarios, where the devices feature a high number of antennas, the channel matrix generation and the computation of the beamforming gain represent up to 90\% of the simulation time ~\cite{testolina2020scalable}. 

In light of these limitations, as the first of our contributions, we optimized the implementation of the 3GPP TR 38.901 model in ns-3 introduced in~\cite{tommaso:20}. 
First, we observed that, as of ns-3.37, part of the trigonometric operations  of the \texttt{GetNewChannel} method of the \texttt{Three\-Gpp\-Channel\-Model} class are unnecessarily repeated for each pair of transmitting and receiving radiating elements. This represents a significant inefficiency, since the inputs of these functions, i.e., the angular parameters of the propagation clusters, depend on the cluster index only. Moreover, the standard library \texttt{sin} and \texttt{cos} functions are particularly demanding to evaluate. Therefore, we cached the trigonometric evaluations of these terms prior to the computation of $\bm{H}$'s coefficients, effectively reducing the complexity of the trigonometric operations from $\mathcal{O}(U \times S \times N)$ to $\mathcal{O}(N)$, where $N$ is the number of propagation clusters. 

Then, we focused on improving the algebra manipulations of the channel matrix performed in the \texttt{Three\-Gpp\-Spectrum\-Propagation\-Loss\-Model} by introducing the support for the open-source library \texttt{Eigen} in ns-3. \texttt{Eigen} is a  linear algebra C\texttt{++} template library that offers fast routines for algebra primitives such as matrix multiplication, decomposition and space transformation~\cite{eigenweb}, and is used by many open-source frameworks such as TensorFlow. 

We set \texttt{Eigen} as an optional, external ns-3 dependency, with the goal of minimizing future code maintenance efforts, 
and thus mimicking the support for other third-party libraries. To get \texttt{Eigen}, ns-3 users can either rely on packet managers, i.e., install the package \texttt{libeigen3-dev} (\texttt{eigen}) for Linux (Mac) systems, or manually install the library by following the official instructions\footnote{\url{https://gitlab.com/libeigen/eigen/-/blob/master/INSTALL}}. Then, \texttt{Eigen} can be enabled via a custom flag defined in the \texttt{macros-\-and-\-definitions.cmake} file, and its presence in the system is shown to the user by exposing whether it has been found or not via the \texttt{ns3-\-config-\-table.cmake} file. The latter also defines the preprocessor definition \texttt{HAVE\_EIGEN3}, which is used in the ns-3 source files to discern \texttt{Eigen}'s availability. Finally, the linking of \texttt{Eigen} with the ns-3 source files is taken care of by the \texttt{CMake} configuration file provided by the library itself, as suggested in the related ns-3 guide.

To prevent the need for \texttt{Eigen} to be installed in the host system, we developed a common set of APIs between the \texttt{Eigen}- and the \gls{stl}-based data structures and primitives. Thanks to this choice, the remainder of the \texttt{spectrum} code is completely abstracted with respect to the presence of the library.
Given that most of the needed operators can not be overloaded for \gls{stl} C\texttt{++} vectors (for instance, \texttt{operator()}), the common interface for both Eigen and \gls{stl}’s based vectors and matrices has been implemented by defining ad hoc structs with custom operators. In particular, we defined:

\begin{itemize}
    \item The complex vector type \texttt{PhasedArrayModel::\-Complex\-Vec\-tor}. This data-structure is defined as an \texttt{std::\-vector} of \texttt{std::\-complex<double>} whenever \texttt{Eigen} is not installed, and as an \texttt{Eigen} vector of \texttt{std::\-complex<double>} otherwise. The set of APIs includes operators \texttt{[]} and \texttt{!=}, which can be used to access the vector entries and to compare pairs of vectors, respectively. Additionally, we defined the \gls{stl}-like methods \texttt{size}, \texttt{norm} and \texttt{resize}, which return the vector size, its $\mathcal{L}^2$-norm, and allow the user to resize the underlying container, respectively. These definitions follow the typical \gls{stl} notation, as it is supported by \texttt{Eigen} as well.
    
    \item The complex matrix type \texttt{MatrixBasedChannelModel::\-Com\-plex\-2DVector}. In this case, the underlying type is a nested \texttt{std::vector} of \texttt{std::complex\-<double>} for when \texttt{Eigen} is disabled, and an \texttt{Eigen} matrix whose entries are of type \texttt{std::\-complex\-<double>} otherwise. 
    
    In this case, we aligned the notation to the APIs provided by \texttt{Eigen}. Specifically, the matrix elements can be accessed via the operator \texttt{()}, which takes as arguments the row and column indices of the entry, while the method \texttt{resize} allows users to resize matrices by specifying the number of rows and columns. In turn, these can be accessed via the \texttt{rows} and \texttt{columns} methods, respectively.

    \item The 3D matrix \texttt{MatrixBased\-ChannelModel::\-Complex\-3D\-Vec\-tor}. This data structure is defined, regardless of \texttt{Eigen}'s availability, as an \texttt{std::\-vector} of \texttt{MatrixBased\-Channel\-Model::\-Complex2DVector}. In this case, the only method provided is \texttt{Multiply\-MatBy\-Left\-And\-RightVec}, which computes a product of the type $\bm{w}_T \bm{H} \bm{w}_R^T$, where $\bm{H} \in \mathbb{C}^{U \times S}$, $\bm{w}_T \in \mathbb{C}^{1 \times U}$ and $\bm{w}_R \in \mathbb{C}^{1 \times S}$. Notably, this computationally demanding evaluation, which is required for computing the beamforming gain in \texttt{Three\-Gpp\-Spectrum\-Propagation\-Loss\-Model}, leverages \texttt{Eigen}'s optimized algorithms whenever the library is installed in the host system.
\end{itemize}

Finally, we remark that the support for \texttt{Eigen} in the ns-3 codebase can possibly be further extended to improve the efficiency of other linear algebra operations, such as the \gls{svd} which is used in the \texttt{mmwave} and \texttt{nr} modules to compute optimal beamformers, and the matrix-by-matrix multiplications needed for relayed channels \cite{9810370}.

\section{A performance-oriented MIMO statistical channel model}
\label{sec:opt_design}

The second approach to reduce computational complexity we propose in this paper is a 
\gls{mimo} channel model for simulating large \gls{m-mimo} scenarios, implemented in the class \texttt{Two\-Ray\-Spectrum\-Propagation\-Loss\-Model}. The goal of this auxiliary model is to offer a faster, albeit slightly less accurate, statistical channel model than the 3GPP TR 38.901 framework of \cite{tommaso:20} by preventing the need for the computation of the complete channel matrix. In line with~\cite{TR38901}, the frequency range of applicability of this model is $0.5 - 100$ GHz, although the framework can be possibly extended to support higher frequencies as well.

The overall channel model design follows the approach of \cite{8445856}, i.e., the end-to-end channel gain is computed by combining several loss and gain terms which account for both large- and small-scale propagation phenomena, and the antenna and beamforming gains.
In particular, let $T$ be a device transmitting a signal $x$ with power $\mathrm{P}_T^x$, and $R$ be another device in the simulation (which may or may not be the intended destination of
$x$). 
The proposed model implements the \texttt{Phased\-Array\-Spectrum\-Propagation\-Loss\-Model} interface by estimating $\mathrm{P}_R^x$, i.e., the power of $x$ received at $R$, as follows:
\begin{align}
\mathrm{P}_R^x[d B m] =& \,\, \mathrm{P}_T^x[d B m] - \mathrm{PL}_{T, R}[d B] \\
        &+ \mathrm{S}_{T, R}[d B] + G_{T, R}[d B] + F_{T, R}[d B], \nonumber
\end{align}
where the terms $\mathrm{PL}_{T, R}$ and $\mathrm{S}_{T, R}$ represent the path loss and the shadowing, respectively, while $G_{T, R}$ and $F_{T, R}$ denote the antenna and beamforming gain and the small-scale fading, respectively. The remainder of this section describes in detail how each of these terms is computed.

\subsection{Path loss, Shadowing, and \gls{los} Condition}
The large-scale propagation phenomena are modeled according to the 3GPP TR 38.901 model~\cite{TR38901}, since its implementation of~\cite{tommaso:20} is not computationally demanding. Nevertheless, the channel model can in principle be coupled with arbitrary classes extending the \texttt{Channel\-Condition\-Model} interface.

Specifically, we first determine the 3GPP scenario. Then, for each link we set the \gls{los} condition in a stochastic manner, using the class extending \texttt{Three\-Gpp\-Channel\-Condition\-Model} which corresponds to the chosen scenario.

Then, we compute the path loss using the 3GPP TR 38.901 formula
\begin{equation}
    PL_{T, R} = A \log_{10} (d) + B + C \log_{10} (f_C) [dB],
\end{equation}
where $d$ is the 3D distance between the transmitter and the receiver, $f_C$ is the carrier frequency, and  $A, B$ and $C$ are model parameters which depend on the specific scenario and the \gls{los} condition.

To account for the presence of blockages, an optional log-normal shadowing component $S_{T, R}$ and an outdoor-to-indoor penetration loss term are added to $PL_{T, R}$.

\subsection{Antenna and Beamforming Gain}


The combined array and beamforming gain is computed using the approach of~\cite{8422746}. 
The proposed model supports the presence of multiple antenna elements at the transmitter and at the receiver, and arbitrary analog beamforming vectors and antenna radiation patterns. Therefore, ns-3 users can use this model in conjunction with any class that implements the \texttt{AntennaModel} interface.
In this implementation, we focus on \glspl{upa}, although the methodology is general and can be applied to arbitrary antenna arrays.

Let $\theta$ and $\varphi$ be the relative zenith and azimuth angles between transmitter and receiver, respectively, and let $\bm{w}\left(\theta_0, \varphi_0\right)$ denote the beamforming vector pointing towards the steering direction $\left(\theta_0, \varphi_0\right)$. We denote with $U = U_h U_v $ the total, horizontal, and vertical number of antenna elements, respectively, and with $ d_h, d_v $ their spacing in the horizontal and vertical domains of the array, respectively. 

Considering first isotropic antennas, the gain pattern of a \gls{upa}, in terms of received power relative to a single radiating element, can be expressed as~\cite{ASPLUND202089}
\begin{equation}
  G_{T, R}^{iso}(\theta, \varphi) = \left| \bm{a_i}^{\mathrm{T}}(\theta, \varphi)  \bm{w}\left(\theta_0, \varphi_0\right) \right|^2,
\end{equation}
where $\bm{a_i}(\theta, \varphi)$ is the array response vector, whose generic entry $m,n$ with $m \in \{0, \ldots, U_v - 1 \}, n \in \{0, \ldots, U_h - 1 \}$ reads
\begin{align}
  a_i (\theta, \varphi)_{m, n} = & \exp \left( j\frac{2\pi}{\lambda}m d_v \cos(\theta) \right) \times \\
    & \exp\left( j \frac{2\pi}{\lambda} n d_h \sin(\theta)\sin(\varphi) \right). \nonumber
\end{align} 

\begin{figure}
  \centering
  \begin{subfigure}[t]{\columnwidth}
    \centering 
    \setlength\fwidth{0.8\columnwidth}
    \setlength\fheight{0.28\columnwidth}
    \input{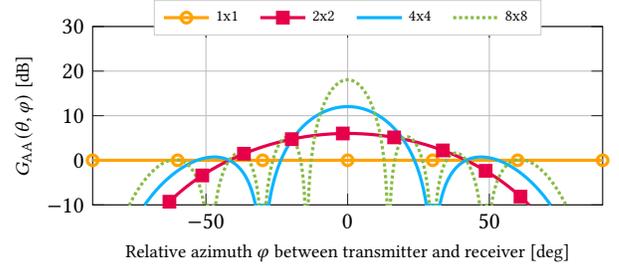}
    \caption{Isotropic Radiating Elements}
    \vspace*{2mm}
    \label{fig:pattern_iso}
  \end{subfigure}
 \hfill
  \begin{subfigure}[t]{\columnwidth}
    \centering 
    \setlength\fwidth{0.8\columnwidth}
    \setlength\fheight{0.28\columnwidth}
    \input{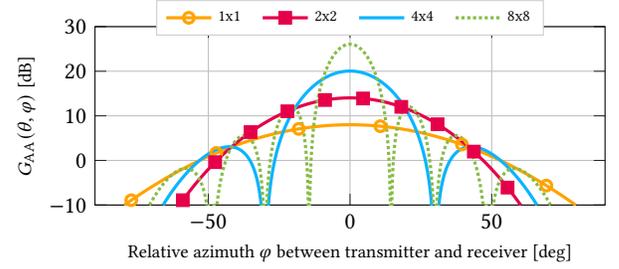}
    \caption{Directional Antenna Radiation Pattern of~\cite[Section~7.3]{TR38901}}
    \label{fig:pattern_3gpp}
  \end{subfigure}
  \caption{Overall Array and Beamforming Gain of a Uniform Planar Array, for Isotropic and 3GPP~\cite[Section~7.3]{TR38901} Radiating Elements and \{1x1, 2x2, 4x4, 8x8\} \glspl{upa}. The Steering Direction is Fixed to $\left(\theta_0, \varphi_0\right) = (0^{\circ}, 0^{\circ})$, and $\theta \equiv 0^{\circ}$}
  \label{Fig:rad_pattern}
\end{figure}

In this work, which supports arbitrary antennas, each antenna element $(m, n)$ actually exhibits a generic radiation pattern $g(\theta, \varphi)_{m, n}$ towards direction $(\theta, \varphi)$. In particular, we assume that $g(\theta, \varphi)_{m, n}$ is constant for all the elements of the array, i.e., $g(\theta, \varphi)_{m, n} \equiv g(\theta, \varphi)$. 
Accordingly, we compute $G_{T, R}(\theta, \varphi)$ in the \texttt{Compute\-Beamforming\-Gain} function of the \texttt{Two\-Ray\-Spectrum\-Propagation\-Loss\-Model} class as
\begin{equation}
  G_{T, R}(\theta, \varphi) =  G_{T, R}^{iso}(\theta, \varphi) \left| g(\theta, \varphi) \right|^2.
\end{equation}
Figures~\ref{fig:pattern_iso} and~\ref{fig:pattern_3gpp} report $G_{T, R} (\theta, \varphi)$ for both the isotropic (\texttt{Isotropic\-Antenna\-Model}) and the 3GPP (\texttt{ThreeGpp\-Antenna\-Model}) radiation patterns, respectively. 

It can be noted that our model abstracts the computation of the received signal power as a \gls{siso} keyhole channel~\cite{chizhik2000capacities}, which is then combined with the spatial antenna gain
patterns at the transmitter/receiver to obtain the received power. This approximation is possibly imprecise when considering \gls{nlos} links, due to the lack of a dominant multipath component. To account for this limitation, we introduce a multiplicative correction factor $\eta$ which scales the beamforming gain as $G^{'}_{T, R}(\theta, \varphi) \equiv \eta G_{T, R}(\theta, \varphi) $. In line with~\cite{kulkarni2018correction}, we set $\eta = 1 / 19$.

\subsection{Fast Fading}

The widely used Rayleigh and Rician distributions fail, even in their generalized forms, to capture the intrinsic bimodality exhibited by \gls{mmwave} scenarios~\cite{yacoub2007kappa, cotton2014human, mavridis2015near}.
Therefore, in our implementation we model fast fading using the more general \gls{ftr} model of~\cite{7917287}. 
This fading model assumes that the received signal comprises two dominant specular components and a mixture of scattered paths, thus modeling the amplitude of the received signal $V_r$ as
\begin{equation}
   V_r = \sqrt{\xi} \exp(j \phi_1) + \sqrt{\xi} \exp(j \phi_2) + X + jY,
\end{equation}
where $\phi_1$, $\phi_2$ are statistically independent random phases, distributed as  $\phi_{i} \sim \mathcal{U} \left[ 0, 2\pi\right]$. $X$ and $Y$ are independent Gaussian random variables, i.e., $X, Y \sim \mathcal{N} (0, \sigma^2)$, which represent the diffuse component of the received signal, which is assumed to be the superposition of multiple weak scattered waves with independent phase. Finally, $\xi$ is a unit-mean Gamma distributed random variable with rate $m$ and \gls{pdf}
\begin{equation}
   f_{\xi} (u) = \frac{m^m u^{m-1}}{\Gamma (m)} exp(-m u).
\end{equation}
In our implementation, $F_{T, R} = \left| V_r \right|^2$ is sampled via the \texttt{Get\-Ftr\-Fast\-Fading} function of the \texttt{Two\-Ray\-Spectrum\-Propagation\-Loss\-Model} class.

The \gls{ftr} fading model is usually expressed as a function of the Gamma rate $m$ and the auxiliary parameters
\begin{align}
    K &\doteq \frac{V_1^2 + V_2^2}{2 \sigma^2} \\
    \Delta &\doteq \frac{2 V_1 V_2}{V_1^2 + V_2^2} \in \left[ 0, 1 \right],
\end{align}
where $K$ represents the ratio of the power of the specular components with respect to the diffuse ones, while $\Delta$ denotes how similar the received powers of the specular components are. By tuning these parameters, a high degree of flexibility can be achieved. Notably, a choice of $\Delta = 0$ effectively yields a Rician-distributed signal amplitude~\cite{7917287}.

\subsubsection{Calibration}

\begin{figure}
    \centering 
    \setlength\fwidth{0.95\columnwidth}
    \setlength\fheight{0.25\columnwidth}
    \input{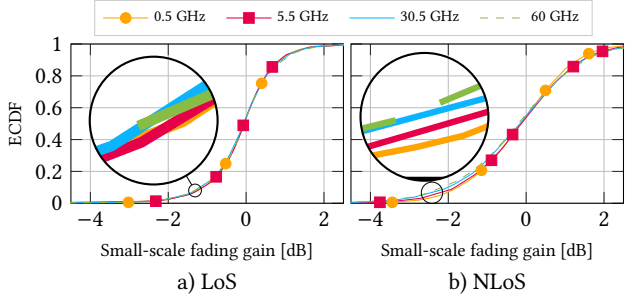}
    \caption{Small-scale Fading Gain Statistics for the UMi Propagation Scenario Versus the Carrier Frequency $f_C$, for both \gls{los} and \gls{nlos} Channel Conditions}
    \label{fig:fading_vs_fc}
\end{figure}

In our work, we calibrated the $V_1, V_2$ and $m$ parameters of the \gls{ftr} fading model using the full 3GPP TR 38.901 channel model as a reference. 
In particular, we first obtained the statistics of the small-scale fading of the 3GPP model, using an ad hoc calibration script (\texttt{three-\-gpp-\-two-\-ray-\-channel-\-calibration.cc}). The script produces a collection of channel gain samples obtained by using the \texttt{Three\-Gpp\-Spectrum\-Propagation\-Loss\-Model} and the \texttt{Three\-Gpp\-Channel\-Model} classes, and neglecting the beamforming gain, path-loss, shadowing and blockages. Accordingly, we isolate the variation around the mean received power caused by the small-scale fading only. 
A separate set of these samples has been retrieved for both \gls{los} and \gls{nlos} channel conditions, the different propagation scenarios of~\cite{TR38901}, and a set of carrier frequencies ranging from $0.5$ to $100$ GHz. However, a preliminary evaluation of the obtained data showed a negligible dependence of the small-scale fading with respect to the carrier frequency, as can be observed in Figure~\ref{fig:fading_vs_fc}. Therefore, we calibrated the \gls{ftr} parameters considering only the channel condition and the propagation scenario.

The small-scale fading samples have been used to estimate the $\Delta, K$ and $m$ \gls{ftr} parameters, and then derive analytically the values of $V_1$ and $V_2$ yielding the fading realizations that are the closest (in a goodness-of-fit sense) to the TR 38.901 model.
To this end, we defined a discrete grid of \gls{ftr} parameters, spanning their whole domain, and considered the corresponding set of parameterized \gls{ftr} distributions. To find the best matching one, we measured the distance between each of these distributions and the 3GPP reference curves by using the Anderson-Darling goodness-of-fit test~\cite{anderson1954test}. This test is used to discern whether a sorted collection of $n$ samples $\{Y_{1}, \ldots, Y_n\}$ originates from a specific distribution, by evaluating the test statistic~\cite{anderson1954test}
\begin{equation}
    A^2 = -n -S(\mathcal{F}),
\end{equation}
where
\begin{equation}
    S(\mathcal{F}) = \sum _{i=1}^{n} \frac {2i-1}{n} \left[ \ln (\mathcal{F}(Y_{i}))+\ln \left(1-(\mathcal{F}(Y_{n+1-i})\right)\right], 
\end{equation}
and $\mathcal{F}$ is the \gls{cdf} of the target distribution.
In the standard Anderson-Darling test, $A^2$ is then compared to a pre-defined critical value to validate the hypothesis. Instead, in our work we find the \gls{ftr} distribution $\mathcal{F}_{m, K, \Delta}$ which yields the lowest $S$. 
Specifically, for each combination of propagation scenario, \gls{los} condition and corresponding samples $\{Y_{1}, \ldots, Y_n\}$ we find
\begin{equation}
    \mathcal{F}_{m^*, K^*, \Delta^*} \doteq \argmin_{m, K, \Delta} S (\mathcal{F}_{m, K, \Delta}).
\end{equation}
Finally, we exported the calibrated \gls{ftr} parameters into ns-3, by storing them in \texttt{SIM\_\-PARAMS\_\-TO\_FTR\_\-PARAMS\_\-TABLE}, i.e., an \texttt{std::map} which associates the propagation scenario and condition to the corresponding best fitting \gls{ftr} parameters.
We remark that this calibration process represents a pre-computation step which needs to be done only once. Indeed, when running a simulation with this channel model, the \gls{ftr} parameters get simply retrieved from the pre-computed lookup table by the \texttt{Get\-Ftr\-Parameters} function. Nevertheless, for the sake of reproducibility and maintainability of the code, we provide this functionality in the Python script \texttt{two-\-ray-\-to-\-three-\-gpp-\-ch-\-calibration.py}.

\section{Benchmarks, examples and use cases}
\label{sec:results}

In this section, we provide an example on how to use the performance-oriented channel model presented above, in conjunction with the \gls{nr}~\cite{patriciello2019e2e} module, to simulate \gls{5g} \gls{mimo} networks. Moreover, we present benchmarks which quantify the simulation time reduction achieved with this work, and we outline some possible use cases.

\subsection{Examples and Benchmarks}

We demonstrate how to use the performance-oriented channel model in the \texttt{cttc-nr-demo-two-ray} script, i.e., a custom version of the \texttt{cttc-nr-demo} example which is included in the \gls{nr} module.
The script deploys $N_{gNB}$ \gls{5g} \gls{nr} base stations, along with $N_{UE}$ users in each cell. Each \gls{ue} uploads data using two \glspl{bwp} operating at 28 and 30~GHz, respectively. Both base stations and user terminals feature \glspl{upa} with multiple radiating elements.

Most simulation parameters can be tuned by ns-3 users. Notably, the script provides the possibility to choose whether to use the 3GPP TR 38.901 channel model of~\cite{tommaso:20} or the \gls{ftr}-based channel model proposed in this work.
In such regard, the use of the \texttt{Two\-Ray\-Spectrum\-Propagation\-Loss\-Model}, instead of the TR 38.901 one, is achieved by:
\begin{enumerate}
\item Setting the \texttt{TypeId} of the \texttt{Spectrum\-Propagation\-Loss\-Model} factory to \texttt{Two\-Ray\-Spectrum\-Propagation\-Loss\-Model}; 
\item Creating an instance of the \texttt{Two\-Ray\-Spectrum\-Propagation\-Loss\-Model} class using the above factory, and setting the corresponding pointer as the \texttt{Spectrum\-Propagation\-Loss\-Model} of both \glspl{bwp}; 
\item  Setting the attribute \texttt{Frequency} of the \texttt{Two\-Ray\-Spectrum\-Propagation\-Loss\-Model} instance as the \gls{bwp} carrier frequency; 
\item  Specifying the 3GPP propagation scenario by setting the attribute \texttt{Scenario}; and 
\item  Creating and setting the \texttt{Channel\-Condition\-Model} by using the \texttt{Two\-Ray\-Spectrum\-Propagation\-Loss\-Model} class \\ \texttt{ChannelConditionModel} attribute. 
\end{enumerate}

On the other hand, the \texttt{Eigen} optimizations simply require users to have the corresponding library installed in their system, and to enable \texttt{Eigen} when configuring ns-3, using the flag \texttt{enable\--eigen}.

\begin{figure}
    \centering 
    \setlength\fwidth{0.95\columnwidth}
    \setlength\fheight{0.3\columnwidth}
    \begin{tikzpicture}

\definecolor{plotColor1}{HTML}{e60049}
\definecolor{plotColor2}{HTML}{0bb4ff}
\definecolor{plotColor3}{HTML}{87bc45}
\definecolor{plotColor4}{HTML}{ffa300}

 \begin{groupplot}[
  group style={
    group size=2 by 1,
    group name=plots,
    horizontal sep= 0.1 cm
  },
    width=0.45\fwidth,
    height=\fheight,
    at={(0\fwidth,0\fheight)},
    scale only axis,
    legend image post style={mark indices={}},
    legend style={
        /tikz/every even column/.append style={column sep=0.2cm},
        at={(1, 1.05)}, 
        anchor=south, 
        draw=white!80!black, 
        font=\scriptsize
        },
    legend columns=4,
    xlabel style={font=\footnotesize},
    xmajorgrids,
    xtick style={color=white!15!black},
    ylabel shift = -1 pt,
    ylabel style={font=\footnotesize},
    ymajorgrids,
    ybar,
    /pgf/bar width=0.6cm,
    ymin=0, ymax=0.8,
    ytick style={color=white!15!black},
    yticklabels=\empty
]

\nextgroupplot[%
               every axis title/.style={below, at={(0.5, -0.38)}},
               ylabel={$T_{A}^{3GPP} / T_{B}$}, 
               yticklabels={0, 0.2, 0.4, 0.6, 0.8, 1},
               ytick={0, 0.2, 0.4, 0.6, 0.8, 1},
               xticklabels={4, 16, 64, 256},
               xtick={1, 2, 3, 4},
               xlabel={Number of antennas at the gNB},
               xmin=0.5, xmax=4.5,]

\addplot[fill=plotColor1] coordinates {
    (1, 0.544278860092163) 
    (2, 0.444827079772949) 
    (3, 0.292946577072144) 
    (4, 0.179594874382019)
};

\nextgroupplot[%
               every axis title/.style={below, at={(0.5, -0.38)}},
               xlabel={Number of UEs},
               xtick={1, 2, 3},
               xticklabels={2, 4, 8},
               xmin=0.5, xmax=3.5,]

\addplot[fill=plotColor2] coordinates {
    (1, 0.67531681060791) 
    (2, 0.54543137550354) 
    (3, 0.514547348022461) 
};

\end{groupplot}
\end{tikzpicture}
    \vspace*{-0.2cm}
    \caption{Ratio of the Median Simulation Times After the Merge of this Work with the Eigen Integration ($T_{A}^{3GPP}$) and as per ns-3.37 ($T_{B}$), when Using the 3GPP Channel Model of~\cite{TR38901}}
    \label{fig:bench_eigen}
\end{figure}
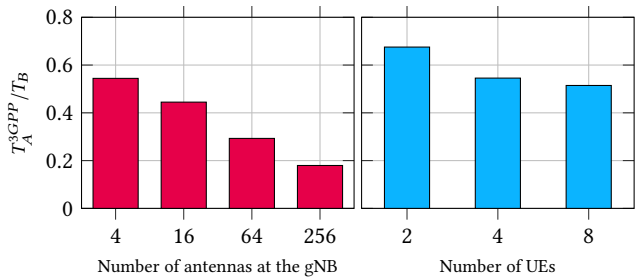

\begin{figure}
    \centering 
    \setlength\fwidth{0.95\columnwidth}
    \setlength\fheight{0.27\columnwidth}
    \begin{tikzpicture}

\definecolor{plotColor1}{HTML}{e60049}
\definecolor{plotColor2}{HTML}{0bb4ff}
\definecolor{plotColor3}{HTML}{87bc45}
\definecolor{plotColor4}{HTML}{ffa300}

 \begin{groupplot}[
  group style={
    group size=2 by 1,
    group name=plots,
    horizontal sep= 0.1 cm
  },
    width=0.45\fwidth,
    height=\fheight,
    at={(0\fwidth,0\fheight)},
    scale only axis,
    legend image post style={mark indices={}},
    legend style={
        /tikz/every even column/.append style={column sep=0.2cm},
        at={(1, 1.05)}, 
        anchor=south, 
        draw=white!80!black, 
        font=\scriptsize
        },
    legend columns=4,
    xlabel style={font=\footnotesize},
    xmajorgrids,
    xtick style={color=white!15!black},
    ylabel shift = -1 pt,
    ylabel style={font=\footnotesize},
    ymajorgrids,
    ybar,
    /pgf/bar width=0.6cm,
    ymin=0, ymax=0.8,
    ytick style={color=white!15!black},
    yticklabels=\empty
]

\nextgroupplot[%
               every axis title/.style={below, at={(0.5, -0.38)}},
               ylabel={$T_{A}^{TR} / T_{B}$}, 
               yticklabels={0, 0.2, 0.4, 0.6, 0.8, 1},
               ytick={0, 0.2, 0.4, 0.6, 0.8, 1},
               xticklabels={4, 16, 64, 256},
               xtick={1, 2, 3, 4},
               xlabel={Number of antennas at the gNB},
               xmin=0.5, xmax=4.5,]

\addplot[fill=plotColor1] coordinates {
    (1, 0.408498287200928) 
    (2, 0.233031988143921) 
    (3, 0.114968657493591) 
    (4, 0.0595511198043823)
};

\nextgroupplot[%
               every axis title/.style={below, at={(0.5, -0.38)}},
               xlabel={Number of UEs},
               xtick={1, 2, 3},
               xticklabels={2, 4, 8},
               xmin=0.5, xmax=3.5,]

\addplot[fill=plotColor2] coordinates {
    (1, 0.408498287200928) 
    (2, 0.296834230422974) 
    (3, 0.142506241798401) 
};

\end{groupplot}
\end{tikzpicture}
    \vspace*{-0.2cm}
    \caption{Ratio of the Median Simulation Times Using the Performance-Oriented Channel Model Presented in this Work ($T_{A}^{TR}$) and the 3GPP Channel Model of~\cite{TR38901} After the Merge of this Work. In this Case, \texttt{Eigen} is Disabled}
    \label{fig:bench_tworay}
\end{figure}
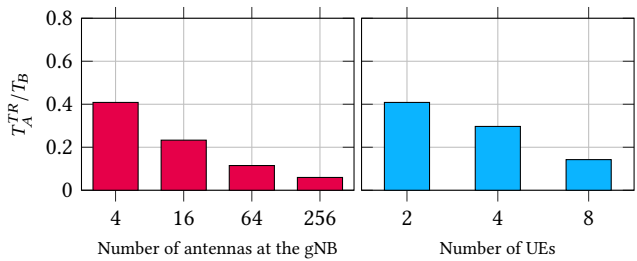
 
We validated our contributions by benchmarking the simulation times exhibited by the above simulation script, which depicts a typical \gls{mimo} \gls{5g} \gls{nr} scenario. To such end, we varied the number of \gls{gnb} antennas and \glspl{ue} deployed, and we timed $100$ simulation runs for each parameter combination. Figure~\ref{fig:bench_eigen} reports the ratio of the median simulation time achieved when using the \texttt{Eigen}-based optimizations, and of the same metric obtained using the vanilla ns-3.37. It can be seen that the matrix multiplication routines offered by \texttt{Eigen} can significantly reduce simulation times. For instance, a reduction of $5$ times in the simulation time is achieved when equipping \glspl{gnb} with $256$ radiating elements.
Similarly, Figure~\ref{fig:bench_tworay} depicts the ratio of the median simulation time obtained by using the \gls{ftr}-based channel model, and the 3GPP TR 38.901 with \texttt{Eigen} disabled. In this case the computational complexity improvement is even more dramatic, with simulations taking as low as 6~\% of the time to complete, with respect to the 3GPP model implementation of~\cite{tommaso:20}. As a reference, the median simulation time obtained on an Intel\textsuperscript{\textcopyright} i5-6700 processor system, before the merge of this work and for $\{2, 4, 8\}$ users is $\{64.7, 210.5, 666.6\}$~[s], respectively.

\begin{figure}
    \centering 
    \setlength\fwidth{0.95\columnwidth}
    \setlength\fheight{0.27\columnwidth}
    \input{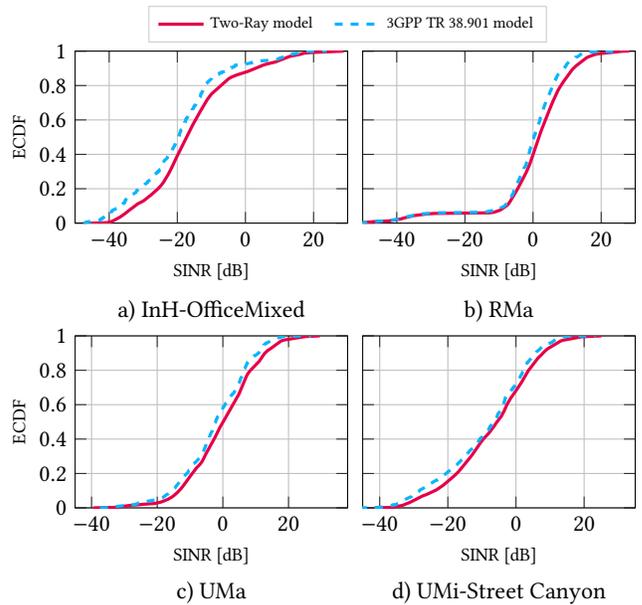}
    \caption{ECDF of the \gls{sinr} Obtained Using the 3GPP Channel Model of~\cite{TR38901}, and the Performance-Oriented Channel Model Presented in this Work, for Different Propagation Scenarios}
    \label{fig:accuracy}
\end{figure}

Finally, we also computed (using the same simulation script, i.e.,~\texttt{cttc-\-nr-\-demo-\-two-\-ray}) the \gls{sinr} statistics achieved by the proposed \gls{ftr}-based model, and compared them to those obtained using the model of~\cite{tommaso:20}. As can be seen in Figure~\ref{fig:accuracy}, the two models provide similar results. Indeed, a non-negligible difference can be found only in the case of the \texttt{InH-OfficeMixed} propagation scenario.

We remark that all the results presented in this section can be reproduced by using the SEM~\cite{magrin2019simulation} scripts which we provide\footnote{\url{https://gitlab.com/pagmatt/ns-3-dev/-/tree/gsoc-wns3}}.

\subsection{Use Cases}

The main goal of both the performance oriented channel model and the optimizations to the 3GPP TR 38.901 model is to enable system-level
simulations of large-scale \gls{mimo} scenarios for which the implementation of~\cite{tommaso:20} exhibits prohibitive computational complexity. Specifically, our contributions allow ns-3 users to simulate wireless deployments where the devices feature antenna arrays with more than hundreds of radiating elements, and/or the number of communication endpoints is particularly high. For example, the modifications presented in this work can be used in the~\gls{nr} and \texttt{mmwave}~\cite{mezzavilla2018end} modules (which both already support the proposed channel models) to simulate massive MIMO 5G NR networks. 
Notably, a preliminary version of the \texttt{Eigen} port has been used in conjunction with the \texttt{mmwave}~\cite{mezzavilla2018end} module to simulate \gls{5g} networks aided by \glspl{irs}, i.e., devices which feature up to $100 \times 100$ reflecting elements~\cite{pagin2022}. 

Moreover, since the supported frequency range is $0.5 - 100$~GHz, this encompasses not only terrestrial \gls{5g} and \gls{lte} deployments, but also most non-terrestrial networks and IEEE \glspl{rat}. Finally, the proposed \texttt{Two\-Ray\-Spectrum\-Propagation\-Loss\-Model} can be further extended to support frequencies above $100$~GHz using reference fading and path loss statistics.

\section{Conclusions and future work}
\label{sec:conc}

In this paper, we presented a set of optimizations concerning the simulation of \gls{mimo} wireless channels in ns-3. First, we introduced the support for the linear algebra library \texttt{Eigen} in ns-3, and reduced the computational complexity of the channel matrix generation procedure by avoiding the unnecessary repetition of trigonometric evaluations. Then, we designed and implemented in ns-3 a performance-oriented statistical channel model based on the \gls{ftr} fading model, which further reduces the simulation time of \gls{mimo} scenarios. 

Profiling results showed that, thanks to this work, the simulation of \gls{mimo} deployments in ns-3 using the 3GPP TR 38.901 channel model takes as little as $20$\% of the original time. Furthermore, whenever the complexity of the simulations represents a major bottleneck, ns-3 users are now given the possibility of using an additional auxiliary channel model, which achieves a further reduction in simulation time, at the cost of a negligible accuracy penalty with respect to the full 3GPP TR 38.901 model.

As part of our future work, we plan to study more refined beamforming gain correction factors, using the 3GPP statistical channel model as a reference, and possibly making the estimation of such term scenario-dependent. Moreover, we envision to design more efficient storage/access data structures and linear algebra operations for 3D matrices, by better leveraging \texttt{Eigen} also in this context.
Finally, we will consider using \gls{simd} for speeding up the evaluation of trigonometric functions, and caching the beamforming gain in the \texttt{Two\-Ray\-Spectrum\-Propagation\-Loss\-Model} class to further reduce the simulation time of \gls{mimo} scenarios in ns-3.

\begin{acks}
The work of Matteo Pagin was partially funded by the Google Summer of Code 2022 program. The work of Michele Polese is partially supported by the U.S. NSF Grant CNS 2225590. CTTC authors have received funding from Grant PID2021-126431OB-I00 funded by MCIN/AEI/10.13039/501100011033 and “ERDF A way of making Europe”, and TSI-063000-2021-56/57 6G-BLUR project by the Spanish Government. This work was partially supported by the European Union under the Italian National Recovery and Resilience Plan (NRRP) of NextGenerationEU, partnership on “Telecommunications of the Future” (PE0000001 - program “RESTART”). Moreover, the authors would like to thank Tom Henderson, Eduardo Almeida, and Gabriel Ferreira for their useful suggestions and support during this work.

\end{acks}

\bibliographystyle{ACM-Reference-Format}
\bibliography{bibl}


\begin{thebibliography}{47}


\ifx \showCODEN    \undefined \def \showCODEN     #1{\unskip}     \fi
\ifx \showDOI      \undefined \def \showDOI       #1{#1}\fi
\ifx \showISBNx    \undefined \def \showISBNx     #1{\unskip}     \fi
\ifx \showISBNxiii \undefined \def \showISBNxiii  #1{\unskip}     \fi
\ifx \showISSN     \undefined \def \showISSN      #1{\unskip}     \fi
\ifx \showLCCN     \undefined \def \showLCCN      #1{\unskip}     \fi
\ifx \shownote     \undefined \def \shownote      #1{#1}          \fi
\ifx \showarticletitle \undefined \def \showarticletitle #1{#1}   \fi
\ifx \showURL      \undefined \def \showURL       {\relax}        \fi
\providecommand\bibfield[2]{#2}
\providecommand\bibinfo[2]{#2}
\providecommand\natexlab[1]{#1}
\providecommand\showeprint[2][]{arXiv:#2}

\bibitem[{3GPP}(2019)]%
        {TR38901}
\bibfield{author}{\bibinfo{person}{{3GPP}}.} \bibinfo{year}{2019}\natexlab{}.
\newblock \bibinfo{title}{{Study on Channel Model for Frequencies from 0.5 to
  100 GHz}}.
\newblock \bibinfo{howpublished}{TR 38.901 (Rel. 15), V15.0.0}.
\newblock


\bibitem[Anderson and Darling(1954)]%
        {anderson1954test}
\bibfield{author}{\bibinfo{person}{Theodore~W Anderson} {and}
  \bibinfo{person}{Donald~A Darling}.} \bibinfo{year}{1954}\natexlab{}.
\newblock \showarticletitle{A Test of Goodness of Fit}.
\newblock \bibinfo{journal}{\emph{Journal of the American statistical
  association}} \bibinfo{volume}{49}, \bibinfo{number}{268}
  (\bibinfo{year}{1954}), \bibinfo{pages}{765--769}.
\newblock


\bibitem[Andrews et~al\mbox{.}(2017)]%
        {andrews2017modeling}
\bibfield{author}{\bibinfo{person}{Jeffrey~G. Andrews},
  \bibinfo{person}{Tianyang Bai}, \bibinfo{person}{Mandar N.~Kulkarni},
  \bibinfo{person}{Ahmed Alkhateeb}, \bibinfo{person}{Abhishek~K. Gupta}, {and}
  \bibinfo{person}{Robert W.~Heath}.} \bibinfo{year}{2017}\natexlab{}.
\newblock \showarticletitle{Modeling and Analyzing Millimeter Wave Cellular
  Systems}.
\newblock \bibinfo{journal}{\emph{IEEE Transactions on Communications}}
  \bibinfo{volume}{65}, \bibinfo{number}{1} (\bibinfo{date}{Jan.}
  \bibinfo{year}{2017}), \bibinfo{pages}{403--430}.
\newblock


\bibitem[{Anuraag Bodi, Steve Blandino, Neeraj Varshney, Jiayi Zhang, Tanguy
  Ropitault, Mattia Lecci, Paolo Testolina, Jian Wang, Chiehping Lai, and
  Camillo Gentile}(2021)]%
        {QD}
\bibfield{author}{\bibinfo{person}{{Anuraag Bodi, Steve Blandino, Neeraj
  Varshney, Jiayi Zhang, Tanguy Ropitault, Mattia Lecci, Paolo Testolina, Jian
  Wang, Chiehping Lai, and Camillo Gentile}}.} \bibinfo{year}{2021}\natexlab{}.
\newblock \bibinfo{title}{{The NIST Q-D Channel Realization Software}}.
\newblock
\newblock


\bibitem[Ashtari~Gargari et~al\mbox{.}(2023)]%
        {amir2023safehaul}
\bibfield{author}{\bibinfo{person}{Amir Ashtari~Gargari},
  \bibinfo{person}{Andrea Ortiz}, \bibinfo{person}{Matteo Pagin},
  \bibinfo{person}{Anja Klein}, \bibinfo{person}{Matthias Hollick},
  \bibinfo{person}{Michele Zorzi}, {and} \bibinfo{person}{Arash Asadi}.}
  \bibinfo{year}{2023}\natexlab{}.
\newblock \showarticletitle{{Safehaul: Risk-Averse Learning for Reliable mmWave
  Self-Backhauling in 6G Networks}}. In \bibinfo{booktitle}{\emph{IEEE
  Conference on Computer Communications (INFOCOM)}}. \bibinfo{address}{New
  York, USA}.
\newblock


\bibitem[Asplund et~al\mbox{.}(2020)]%
        {ASPLUND202089}
\bibfield{author}{\bibinfo{person}{Henrik Asplund}, \bibinfo{person}{David
  Astely}, \bibinfo{person}{Peter von Butovitsch}, \bibinfo{person}{Thomas
  Chapman}, \bibinfo{person}{Mattias Frenne}, \bibinfo{person}{Farshid
  Ghasemzadeh}, \bibinfo{person}{Måns Hagström}, \bibinfo{person}{Billy
  Hogan}, \bibinfo{person}{George Jöngren}, \bibinfo{person}{Jonas Karlsson},
  \bibinfo{person}{Fredric Kronestedt}, {and} \bibinfo{person}{Erik Larsson}.}
  \bibinfo{year}{2020}\natexlab{}.
\newblock \showarticletitle{Chapter 4 - Antenna Arrays and Classical
  Beamforming}.
\newblock In \bibinfo{booktitle}{\emph{Advanced Antenna Systems for 5G Network
  Deployments}}. \bibinfo{publisher}{Academic Press}, \bibinfo{pages}{89--132}.
\newblock


\bibitem[Chizhik et~al\mbox{.}(2000)]%
        {chizhik2000capacities}
\bibfield{author}{\bibinfo{person}{Dmitry Chizhik}, \bibinfo{person}{Gerald~J.
  Foschini}, {and} \bibinfo{person}{Reinaldo~A. Valenzuela}.}
  \bibinfo{year}{2000}\natexlab{}.
\newblock \showarticletitle{Capacities of Multi-element Transmit and Receive
  Antennas: Correlations and Keyholes}.
\newblock \bibinfo{journal}{\emph{Electronics Letters}} \bibinfo{volume}{36},
  \bibinfo{number}{13} (\bibinfo{date}{Jun.} \bibinfo{year}{2000}),
  \bibinfo{pages}{1}.
\newblock


\bibitem[Cotton(2014)]%
        {cotton2014human}
\bibfield{author}{\bibinfo{person}{Simon~L Cotton}.}
  \bibinfo{year}{2014}\natexlab{}.
\newblock \showarticletitle{Human Body Shadowing in Cellular Device-to-Device
  Communications: Channel Modeling Using the Shadowed $\kappa$-$\mu$ Fading
  Model}.
\newblock \bibinfo{journal}{\emph{IEEE Journal on Selected areas in
  Communications}} \bibinfo{volume}{33}, \bibinfo{number}{1}
  (\bibinfo{date}{Nov.} \bibinfo{year}{2014}), \bibinfo{pages}{111--119}.
\newblock


\bibitem[Ghosh et~al\mbox{.}(2019)]%
        {ghosh20195g}
\bibfield{author}{\bibinfo{person}{Amitabha Ghosh}, \bibinfo{person}{Andreas
  Maeder}, \bibinfo{person}{Matthew Baker}, {and} \bibinfo{person}{Devaki
  Chandramouli}.} \bibinfo{year}{2019}\natexlab{}.
\newblock \showarticletitle{{5G} Evolution: A View on {5G} Cellular Technology
  Beyond {3GPP} Release 15}.
\newblock \bibinfo{journal}{\emph{IEEE Access}}  \bibinfo{volume}{7}
  (\bibinfo{date}{Sep.} \bibinfo{year}{2019}), \bibinfo{pages}{127639--127651}.
\newblock


\bibitem[Guennebaud et~al\mbox{.}(2010)]%
        {eigenweb}
\bibfield{author}{\bibinfo{person}{Ga\"{e}l Guennebaud},
  \bibinfo{person}{Beno\^{i}t Jacob}, {et~al\mbox{.}}}
  \bibinfo{year}{2010}\natexlab{}.
\newblock \bibinfo{title}{Eigen v3}.
\newblock \bibinfo{howpublished}{http://eigen.tuxfamily.org}.
\newblock


\bibitem[Han and Chen(2018)]%
        {han2018propagation}
\bibfield{author}{\bibinfo{person}{Chong Han} {and} \bibinfo{person}{Yi Chen}.}
  \bibinfo{year}{2018}\natexlab{}.
\newblock \showarticletitle{{Propagation Modeling for Wireless Communications
  in the Terahertz Band}}.
\newblock \bibinfo{journal}{\emph{IEEE Communications Magazine}}
  \bibinfo{volume}{56}, \bibinfo{number}{6} (\bibinfo{date}{Jun.}
  \bibinfo{year}{2018}), \bibinfo{pages}{96--101}.
\newblock
\showISSN{1558-1896}


\bibitem[Hemadeh et~al\mbox{.}(2017)]%
        {hemadeh2018millimeter}
\bibfield{author}{\bibinfo{person}{Ibrahim~A. Hemadeh}, \bibinfo{person}{Katla
  Satyanarayana}, \bibinfo{person}{Mohammed El-Hajjar}, {and}
  \bibinfo{person}{Lajos Hanzo}.} \bibinfo{year}{2017}\natexlab{}.
\newblock \showarticletitle{Millimeter-Wave Communications: Physical Channel
  Models, Design Considerations, Antenna Constructions and Link-Budget}.
\newblock \bibinfo{journal}{\emph{IEEE Communications Surveys Tutorials}}
  \bibinfo{volume}{20}, \bibinfo{number}{99} (\bibinfo{date}{Dec.}
  \bibinfo{year}{2017}), \bibinfo{pages}{870--913}.
\newblock


\bibitem[Heng et~al\mbox{.}(2021)]%
        {heng2021six}
\bibfield{author}{\bibinfo{person}{Yuqiang Heng}, \bibinfo{person}{Jeffrey~G
  Andrews}, \bibinfo{person}{Jianhua Mo}, \bibinfo{person}{Vutha Va},
  \bibinfo{person}{Anum Ali}, \bibinfo{person}{Boon~Loong Ng}, {and}
  \bibinfo{person}{Jianzhong~Charlie Zhang}.} \bibinfo{year}{2021}\natexlab{}.
\newblock \showarticletitle{Six Key Challenges for Beam Management in 5.5G and
  6G Systems}.
\newblock \bibinfo{journal}{\emph{IEEE Communications Magazine}}
  \bibinfo{volume}{59}, \bibinfo{number}{7} (\bibinfo{date}{Jul.}
  \bibinfo{year}{2021}), \bibinfo{pages}{74--79}.
\newblock


\bibitem[Hoydis et~al\mbox{.}(2021)]%
        {hoydis2021toward}
\bibfield{author}{\bibinfo{person}{Jakob Hoydis},
  \bibinfo{person}{Fay{\c{c}}al~Ait Aoudia}, \bibinfo{person}{Alvaro Valcarce},
  {and} \bibinfo{person}{Harish Viswanathan}.} \bibinfo{year}{2021}\natexlab{}.
\newblock \showarticletitle{Toward a {6G} {AI}-Native Air Interface}.
\newblock \bibinfo{journal}{\emph{IEEE Communications Magazine}}
  \bibinfo{volume}{59}, \bibinfo{number}{5} (\bibinfo{date}{May}
  \bibinfo{year}{2021}), \bibinfo{pages}{76--81}.
\newblock


\bibitem[ITU-R(2015)]%
        {itu-r-2083}
\bibfield{author}{\bibinfo{person}{ITU-R}.} \bibinfo{year}{2015}\natexlab{}.
\newblock \bibinfo{booktitle}{\emph{{IMT} Vision - Framework and Overall
  Objectives of the Future Development of {IMT} for 2020 and Beyond}}.
\newblock \bibinfo{type}{Report {M}.2083}.
\newblock


\bibitem[ITU-R(2022)]%
        {imt2030}
\bibfield{author}{\bibinfo{person}{ITU-R}.} \bibinfo{year}{2022}\natexlab{}.
\newblock \bibinfo{booktitle}{\emph{Future Technology Trends of Terrestrial
  International Mobile Telecommunications Systems Towards 2030 and Beyond}}.
\newblock \bibinfo{type}{Report} M.2516-0.
\newblock


\bibitem[Jin et~al\mbox{.}(2021a)]%
        {jin2021eesm}
\bibfield{author}{\bibinfo{person}{Sian Jin}, \bibinfo{person}{Sumit Roy},
  {and} \bibinfo{person}{Thomas~R Henderson}.}
  \bibinfo{year}{2021}\natexlab{a}.
\newblock \showarticletitle{{EESM}-log-{AR}: an Efficient Error Model for {OFDM
  MIMO} Systems Over Time-Varying Channels}. In
  \bibinfo{booktitle}{\emph{Proceedings of the 2021 Workshop on ns-3}}.
  \bibinfo{publisher}{Association for Computing Machinery},
  \bibinfo{address}{Virtual Event, USA}.
\newblock


\bibitem[Jin et~al\mbox{.}(2021b)]%
        {jin2021efficient}
\bibfield{author}{\bibinfo{person}{Sian Jin}, \bibinfo{person}{Sumit Roy},
  {and} \bibinfo{person}{Thomas~R Henderson}.}
  \bibinfo{year}{2021}\natexlab{b}.
\newblock \showarticletitle{Efficient {PHY} Layer Abstraction for Fast
  Simulations in Complex System Environments}.
\newblock \bibinfo{journal}{\emph{IEEE Transactions on Communications}}
  \bibinfo{volume}{69}, \bibinfo{number}{8} (\bibinfo{date}{May}
  \bibinfo{year}{2021}), \bibinfo{pages}{5649--5660}.
\newblock


\bibitem[Jin et~al\mbox{.}(2020)]%
        {jin2020efficient}
\bibfield{author}{\bibinfo{person}{Sian Jin}, \bibinfo{person}{Sumit Roy},
  \bibinfo{person}{Weihua Jiang}, {and} \bibinfo{person}{Thomas~R Henderson}.}
  \bibinfo{year}{2020}\natexlab{}.
\newblock \showarticletitle{Efficient Abstractions for Implementing {TGn}
  Channel and {OFDM-MIMO} Links in ns-3}. In
  \bibinfo{booktitle}{\emph{Proceedings of the 2020 Workshop on ns-3}}.
  \bibinfo{publisher}{Association for Computing Machinery},
  \bibinfo{address}{Gaithersburg, MD, USA}.
\newblock


\bibitem[Jornet and Akyildiz(2011)]%
        {jornet2011channel}
\bibfield{author}{\bibinfo{person}{Josep~Miquel Jornet} {and}
  \bibinfo{person}{Ian~F Akyildiz}.} \bibinfo{year}{2011}\natexlab{}.
\newblock \showarticletitle{Channel Modeling and Capacity Analysis for
  Electromagnetic Wireless Nanonetworks in the Terahertz Band}.
\newblock \bibinfo{journal}{\emph{IEEE Transactions on Wireless
  Communications}} \bibinfo{volume}{10}, \bibinfo{number}{10}
  (\bibinfo{date}{Aug.} \bibinfo{year}{2011}), \bibinfo{pages}{3211--3221}.
\newblock


\bibitem[Kulkarni et~al\mbox{.}(2018)]%
        {kulkarni2018correction}
\bibfield{author}{\bibinfo{person}{Mandar~N Kulkarni}, \bibinfo{person}{Eugene
  Visotsky}, {and} \bibinfo{person}{Jeffrey~G Andrews}.}
  \bibinfo{year}{2018}\natexlab{}.
\newblock \showarticletitle{Correction Factor for Analysis of MIMO Wireless
  Networks with Highly Directional Beamforming}.
\newblock \bibinfo{journal}{\emph{IEEE Wireless Communications Letters}}
  \bibinfo{volume}{7}, \bibinfo{number}{5} (\bibinfo{date}{Mar.}
  \bibinfo{year}{2018}), \bibinfo{pages}{756--759}.
\newblock


\bibitem[Lacava et~al\mbox{.}(2022)]%
        {lacava2022programmable}
\bibfield{author}{\bibinfo{person}{Andrea Lacava}, \bibinfo{person}{Michele
  Polese}, \bibinfo{person}{Rajarajan Sivaraj}, \bibinfo{person}{Rahul
  Soundrarajan}, \bibinfo{person}{Bhawani~Shanker Bhati},
  \bibinfo{person}{Tarunjeet Singh}, \bibinfo{person}{Tommaso Zugno},
  \bibinfo{person}{Francesca Cuomo}, {and} \bibinfo{person}{Tommaso Melodia}.}
  \bibinfo{year}{2022}\natexlab{}.
\newblock \showarticletitle{Programmable and Customized Intelligence for
  Traffic Steering in 5G Networks using Open RAN Architectures}.
\newblock \bibinfo{journal}{\emph{arXiv preprint arXiv:2209.14171}}
  (\bibinfo{year}{2022}).
\newblock


\bibitem[Lagen et~al\mbox{.}(2020)]%
        {lagen2020new}
\bibfield{author}{\bibinfo{person}{Sandra Lagen}, \bibinfo{person}{Kevin
  Wanuga}, \bibinfo{person}{Hussain Elkotby}, \bibinfo{person}{Sanjay Goyal},
  \bibinfo{person}{Natale Patriciello}, {and} \bibinfo{person}{Lorenza
  Giupponi}.} \bibinfo{year}{2020}\natexlab{}.
\newblock \showarticletitle{New Radio Physical Layer Abstraction for
  System-Level Simulations of {5G} Networks}. In
  \bibinfo{booktitle}{\emph{International Conference on Communications (ICC)}}.
  \bibinfo{publisher}{IEEE}, \bibinfo{address}{Virtual Event}.
\newblock


\bibitem[Lecci et~al\mbox{.}(2020)]%
        {lecci2020simplified}
\bibfield{author}{\bibinfo{person}{Mattia Lecci}, \bibinfo{person}{Paolo
  Testolina}, \bibinfo{person}{Marco Giordani}, \bibinfo{person}{Michele
  Polese}, \bibinfo{person}{Tanguy Ropitault}, \bibinfo{person}{Camillo
  Gentile}, \bibinfo{person}{Neeraj Varshney}, \bibinfo{person}{Anuraag Bodi},
  {and} \bibinfo{person}{Michele Zorzi}.} \bibinfo{year}{2020}\natexlab{}.
\newblock \showarticletitle{{Simplified Ray Tracing for the Millimeter Wave
  Channel: A Performance Evaluation}}. In \bibinfo{booktitle}{\emph{Information
  Theory and Applications Workshop (ITA)}}. \bibinfo{publisher}{IEEE},
  \bibinfo{address}{San Diego, CA USA}.
\newblock


\bibitem[Liu et~al\mbox{.}(2021b)]%
        {9444237}
\bibfield{author}{\bibinfo{person}{Shanyun Liu}, \bibinfo{person}{Xianbin Yu},
  \bibinfo{person}{Rongbin Guo}, \bibinfo{person}{Yajie Tang}, {and}
  \bibinfo{person}{Zhifeng Zhao}.} \bibinfo{year}{2021}\natexlab{b}.
\newblock \showarticletitle{THz Channel Modeling: Consolidating the Road to THz
  Communications}.
\newblock \bibinfo{journal}{\emph{China Communications}} \bibinfo{volume}{18},
  \bibinfo{number}{5} (\bibinfo{date}{May} \bibinfo{year}{2021}),
  \bibinfo{pages}{33--49}.
\newblock


\bibitem[Liu et~al\mbox{.}(2021a)]%
        {liu2021performance}
\bibfield{author}{\bibinfo{person}{Yuchen Liu}, \bibinfo{person}{Shelby~K
  Crisp}, {and} \bibinfo{person}{Douglas~M Blough}.}
  \bibinfo{year}{2021}\natexlab{a}.
\newblock \showarticletitle{Performance Study of Statistical and Deterministic
  Channel Models for mmWave Wi-Fi Networks in ns-3}. In
  \bibinfo{booktitle}{\emph{Proceedings of the 2021 Workshop on ns-3}}.
  \bibinfo{publisher}{Association for Computing Machinery},
  \bibinfo{address}{Virtual Event, USA}.
\newblock


\bibitem[Magrin et~al\mbox{.}(2019)]%
        {magrin2019simulation}
\bibfield{author}{\bibinfo{person}{Davide Magrin}, \bibinfo{person}{Dizhi
  Zhou}, {and} \bibinfo{person}{Michele Zorzi}.}
  \bibinfo{year}{2019}\natexlab{}.
\newblock \showarticletitle{A Simulation Execution Manager for ns-3:
  Encouraging Reproducibility and Simplifying Statistical Analysis of ns-3
  Simulations}. In \bibinfo{booktitle}{\emph{Proceedings of the 22nd
  International ACM Conference on Modeling, Analysis and Simulation of Wireless
  and Mobile Systems}}. \bibinfo{address}{Miami Beach, FL USA}.
\newblock


\bibitem[Mavridis et~al\mbox{.}(2015)]%
        {mavridis2015near}
\bibfield{author}{\bibinfo{person}{Theodoros Mavridis}, \bibinfo{person}{Luca
  Petrillo}, \bibinfo{person}{Julien Sarrazin}, \bibinfo{person}{Aziz
  Benlarbi-Delai}, {and} \bibinfo{person}{Philippe De~Doncker}.}
  \bibinfo{year}{2015}\natexlab{}.
\newblock \showarticletitle{Near-Body Shadowing Analysis at 60 GHz}.
\newblock \bibinfo{journal}{\emph{IEEE Transactions on Antennas and
  Propagation}} \bibinfo{volume}{63}, \bibinfo{number}{10}
  (\bibinfo{date}{Jul.} \bibinfo{year}{2015}), \bibinfo{pages}{4505--4511}.
\newblock


\bibitem[Mezzavilla et~al\mbox{.}(2018)]%
        {mezzavilla2018end}
\bibfield{author}{\bibinfo{person}{Marco Mezzavilla}, \bibinfo{person}{Menglei
  Zhang}, \bibinfo{person}{Michele Polese}, \bibinfo{person}{Russell Ford},
  \bibinfo{person}{Sourjya Dutta}, \bibinfo{person}{Sundeep Rangan}, {and}
  \bibinfo{person}{Michele Zorzi}.} \bibinfo{year}{2018}\natexlab{}.
\newblock \showarticletitle{End-to-End Simulation of 5G mmWave Networks}.
\newblock \bibinfo{journal}{\emph{IEEE Communications Surveys \& Tutorials}}
  \bibinfo{volume}{20}, \bibinfo{number}{3} (\bibinfo{date}{Apr.}
  \bibinfo{year}{2018}), \bibinfo{pages}{2237--2263}.
\newblock


\bibitem[Pagin et~al\mbox{.}(2022a)]%
        {9810370}
\bibfield{author}{\bibinfo{person}{Matteo Pagin}, \bibinfo{person}{Marco
  Giordani}, \bibinfo{person}{Amir~Ashtari Gargari}, \bibinfo{person}{Alberto
  Rech}, \bibinfo{person}{Federico Moretto}, \bibinfo{person}{Stefano Tomasin},
  \bibinfo{person}{Jonathan Gambini}, {and} \bibinfo{person}{Michele Zorzi}.}
  \bibinfo{year}{2022}\natexlab{a}.
\newblock \showarticletitle{End-to-End Simulation of {5G} Networks Assisted by
  {IRS} and {AF} Relays}. In \bibinfo{booktitle}{\emph{Proc.\ IEEE MedComNet}}.
  \bibinfo{address}{Paphos, Cyprus}.
\newblock


\bibitem[Pagin et~al\mbox{.}(2022b)]%
        {pagin2022}
\bibfield{author}{\bibinfo{person}{Matteo Pagin}, \bibinfo{person}{Tommaso
  Zugno}, \bibinfo{person}{Michele Polese}, {and} \bibinfo{person}{Michele
  Zorzi}.} \bibinfo{year}{2022}\natexlab{b}.
\newblock \showarticletitle{{Resource Management for 5G NR Integrated Access
  and Backhaul: A Semi-Centralized Approach}}.
\newblock \bibinfo{journal}{\emph{IEEE Transactions on Wireless
  Communications}} \bibinfo{volume}{21}, \bibinfo{number}{2}
  (\bibinfo{date}{Jul.} \bibinfo{year}{2022}), \bibinfo{pages}{753--767}.
\newblock


\bibitem[Patriciello et~al\mbox{.}(2019)]%
        {patriciello2019e2e}
\bibfield{author}{\bibinfo{person}{Natale Patriciello}, \bibinfo{person}{Sandra
  Lagen}, \bibinfo{person}{Biljana Bojovic}, {and} \bibinfo{person}{Lorenza
  Giupponi}.} \bibinfo{year}{2019}\natexlab{}.
\newblock \showarticletitle{An E2E Simulator for 5G NR Networks}.
\newblock \bibinfo{journal}{\emph{Simulation Modelling Practice and Theory}}
  \bibinfo{volume}{96} (\bibinfo{date}{Nov.} \bibinfo{year}{2019}),
  \bibinfo{pages}{101933}.
\newblock


\bibitem[Polese et~al\mbox{.}(2022)]%
        {polese2022colo}
\bibfield{author}{\bibinfo{person}{Michele Polese}, \bibinfo{person}{Leonardo
  Bonati}, \bibinfo{person}{Salvatore D’Oro}, \bibinfo{person}{Stefano
  Basagni}, {and} \bibinfo{person}{Tommaso Melodia}.}
  \bibinfo{year}{2022}\natexlab{}.
\newblock \showarticletitle{ColO-RAN: Developing Machine Learning-Based xApps
  for Open RAN Closed-Loop Control on Programmable Experimental Platforms}.
\newblock \bibinfo{journal}{\emph{IEEE Transactions on Mobile Computing}}
  (\bibinfo{date}{Jul.} \bibinfo{year}{2022}), \bibinfo{pages}{1--14}.
\newblock


\bibitem[Polese et~al\mbox{.}(2023)]%
        {polese2023understanding}
\bibfield{author}{\bibinfo{person}{Michele Polese}, \bibinfo{person}{Leonardo
  Bonati}, \bibinfo{person}{Salvatore D’Oro}, \bibinfo{person}{Stefano
  Basagni}, {and} \bibinfo{person}{Tommaso Melodia}.}
  \bibinfo{year}{2023}\natexlab{}.
\newblock \showarticletitle{Understanding O-RAN: Architecture, Interfaces,
  Algorithms, Security, and Research Challenges}.
\newblock \bibinfo{journal}{\emph{IEEE Communications Surveys \& Tutorials}}
  (\bibinfo{date}{Jan.} \bibinfo{year}{2023}), \bibinfo{pages}{Early Access}.
\newblock


\bibitem[Polese et~al\mbox{.}(2021)]%
        {polese20216g}
\bibfield{author}{\bibinfo{person}{Michele Polese}, \bibinfo{person}{Marco
  Giordani}, \bibinfo{person}{Marco Mezzavilla}, \bibinfo{person}{Sundeep
  Rangan}, {and} \bibinfo{person}{Michele Zorzi}.}
  \bibinfo{year}{2021}\natexlab{}.
\newblock \showarticletitle{6G Enabling Technologies}.
\newblock In \bibinfo{booktitle}{\emph{{6G} mobile wireless networks}}.
  \bibinfo{publisher}{Springer}, \bibinfo{pages}{25--41}.
\newblock


\bibitem[Polese et~al\mbox{.}(2020)]%
        {polese2020toward}
\bibfield{author}{\bibinfo{person}{Michele Polese},
  \bibinfo{person}{Josep~Miquel Jornet}, \bibinfo{person}{Tommaso Melodia},
  {and} \bibinfo{person}{Michele Zorzi}.} \bibinfo{year}{2020}\natexlab{}.
\newblock \showarticletitle{{Toward End-to-End, Full-Stack 6G Terahertz
  Networks}}.
\newblock \bibinfo{journal}{\emph{IEEE Communications Magazine}}
  \bibinfo{volume}{58}, \bibinfo{number}{11} (\bibinfo{date}{Nov.}
  \bibinfo{year}{2020}), \bibinfo{pages}{48--54}.
\newblock


\bibitem[Polese and Zorzi(2018)]%
        {8445856}
\bibfield{author}{\bibinfo{person}{Michele Polese} {and}
  \bibinfo{person}{Michele Zorzi}.} \bibinfo{year}{2018}\natexlab{}.
\newblock \showarticletitle{Impact of Channel Models on the End-to-End
  Performance of Mmwave Cellular Networks}. In \bibinfo{booktitle}{\emph{IEEE
  19th International Workshop on Signal Processing Advances in Wireless
  Communications (SPAWC)}}. \bibinfo{address}{Kalamata, Greece}.
\newblock


\bibitem[Ramos et~al\mbox{.}(2022)]%
        {10.1145/3532577.3532596}
\bibfield{author}{\bibinfo{person}{Andrea Ramos}, \bibinfo{person}{Yanet
  Estrada}, \bibinfo{person}{Miguel Cantero}, \bibinfo{person}{Jaime Romero},
  \bibinfo{person}{David Mart\'{\i}n-Sacrist\'{a}n}, \bibinfo{person}{Sa\'{u}l
  Inca}, \bibinfo{person}{Manuel Fuentes}, {and} \bibinfo{person}{Jos\'{e}
  Monserrat}.} \bibinfo{year}{2022}\natexlab{}.
\newblock \showarticletitle{Implementation and Calibration of the 3GPP
  Industrial Channel Model for ns-3}. In \bibinfo{booktitle}{\emph{Proceedings
  of the 2022 Workshop on ns-3}}. \bibinfo{publisher}{Association for Computing
  Machinery}, \bibinfo{address}{Virtual Event, USA}.
\newblock


\bibitem[Rappaport et~al\mbox{.}(2013)]%
        {rappaport2013millimeter}
\bibfield{author}{\bibinfo{person}{Theodore~S. Rappaport}, \bibinfo{person}{Shu
  Sun}, \bibinfo{person}{Rimma Mayzus}, \bibinfo{person}{Hang Zhao},
  \bibinfo{person}{Yaniv Azar}, \bibinfo{person}{Kevin Wang},
  \bibinfo{person}{George N.~Wong}, \bibinfo{person}{Jocelyn K.~Schulz},
  \bibinfo{person}{Mathew Samimi}, {and} \bibinfo{person}{Felix Gutierrez}.}
  \bibinfo{year}{2013}\natexlab{}.
\newblock \showarticletitle{{Millimeter Wave Mobile Communications for 5G
  Cellular: It Will Work!}}
\newblock \bibinfo{journal}{\emph{IEEE Access}}  \bibinfo{volume}{1}
  (\bibinfo{date}{May} \bibinfo{year}{2013}), \bibinfo{pages}{335--349}.
\newblock
\showISSN{2169-3536}


\bibitem[Rebato et~al\mbox{.}(2018)]%
        {8422746}
\bibfield{author}{\bibinfo{person}{Mattia Rebato}, \bibinfo{person}{Laura
  Resteghini}, \bibinfo{person}{Christian Mazzucco}, {and}
  \bibinfo{person}{Michele Zorzi}.} \bibinfo{year}{2018}\natexlab{}.
\newblock \showarticletitle{{Study of Realistic Antenna Patterns in 5G mmWave
  Cellular Scenarios}}. In \bibinfo{booktitle}{\emph{Proc.\ IEEE ICC}}.
  \bibinfo{address}{Kansas City, MO, USA}.
\newblock


\bibitem[Romero-Jerez et~al\mbox{.}(2017)]%
        {7917287}
\bibfield{author}{\bibinfo{person}{Juan~M. Romero-Jerez},
  \bibinfo{person}{F.~Javier Lopez-Martinez}, \bibinfo{person}{José~F. Paris},
  {and} \bibinfo{person}{Andrea~J. Goldsmith}.}
  \bibinfo{year}{2017}\natexlab{}.
\newblock \showarticletitle{The Fluctuating Two-Ray Fading Model: Statistical
  Characterization and Performance Analysis}.
\newblock \bibinfo{journal}{\emph{IEEE Transactions on Wireless
  Communications}} \bibinfo{volume}{16}, \bibinfo{number}{7}
  (\bibinfo{date}{May} \bibinfo{year}{2017}), \bibinfo{pages}{4420--4432}.
\newblock


\bibitem[Shafi et~al\mbox{.}(2018)]%
        {shafi2018microwave}
\bibfield{author}{\bibinfo{person}{Mansoor Shafi}, \bibinfo{person}{Jianhua
  Zhang}, \bibinfo{person}{Harsh Tataria}, \bibinfo{person}{Andreas~F Molisch},
  \bibinfo{person}{Shu Sun}, \bibinfo{person}{Theodore~S Rappaport},
  \bibinfo{person}{Fredrik Tufvesson}, \bibinfo{person}{Shangbin Wu}, {and}
  \bibinfo{person}{Koshiro Kitao}.} \bibinfo{year}{2018}\natexlab{}.
\newblock \showarticletitle{Microwave vs. Millimeter-Wave Propagation Channels:
  Key Differences and Impact on {5G} Cellular Systems}.
\newblock \bibinfo{journal}{\emph{IEEE Communications Magazine}}
  \bibinfo{volume}{56}, \bibinfo{number}{12} (\bibinfo{date}{Dec.}
  \bibinfo{year}{2018}), \bibinfo{pages}{14--20}.
\newblock


\bibitem[Testolina et~al\mbox{.}(2020)]%
        {testolina2020scalable}
\bibfield{author}{\bibinfo{person}{Paolo Testolina}, \bibinfo{person}{Mattia
  Lecci}, \bibinfo{person}{Michele Polese}, \bibinfo{person}{Marco Giordani},
  {and} \bibinfo{person}{Michele Zorzi}.} \bibinfo{year}{2020}\natexlab{}.
\newblock \showarticletitle{Scalable and Accurate Modeling of the Millimeter
  Wave Channel}. In \bibinfo{booktitle}{\emph{International Conference on
  Computing, Networking and Communications (ICNC)}}. IEEE,
  \bibinfo{address}{Big Island, Hawaii, USA}.
\newblock


\bibitem[Yacoub(2007)]%
        {yacoub2007kappa}
\bibfield{author}{\bibinfo{person}{Michel~Daoud Yacoub}.}
  \bibinfo{year}{2007}\natexlab{}.
\newblock \showarticletitle{The $\kappa$-$\mu$ Distribution and the
  $\eta$-$\mu$ Distribution}.
\newblock \bibinfo{journal}{\emph{IEEE Antennas and Propagation Magazine}}
  \bibinfo{volume}{49}, \bibinfo{number}{1} (\bibinfo{date}{Feb.}
  \bibinfo{year}{2007}), \bibinfo{pages}{68--81}.
\newblock


\bibitem[Zhang et~al\mbox{.}(2019)]%
        {zhang2019will}
\bibfield{author}{\bibinfo{person}{Menglei Zhang}, \bibinfo{person}{Michele
  Polese}, \bibinfo{person}{Marco Mezzavilla}, \bibinfo{person}{Jing Zhu},
  \bibinfo{person}{Sundeep Rangan}, \bibinfo{person}{Shivendra Panwar}, {and}
  \bibinfo{person}{Michele Zorzi}.} \bibinfo{year}{2019}\natexlab{}.
\newblock \showarticletitle{{Will TCP Work in mmWave 5G Cellular Networks?}}
\newblock \bibinfo{journal}{\emph{IEEE Communications Magazine}}
  \bibinfo{volume}{57}, \bibinfo{number}{1} (\bibinfo{date}{Jan.}
  \bibinfo{year}{2019}), \bibinfo{pages}{65--71}.
\newblock
\showISSN{1558-1896}


\bibitem[Zugno et~al\mbox{.}(2021)]%
        {10.1145/3460797.3460801}
\bibfield{author}{\bibinfo{person}{Tommaso Zugno}, \bibinfo{person}{Matteo
  Drago}, \bibinfo{person}{Sandra Lag\'{e}n}, \bibinfo{person}{Zoraze Ali},
  {and} \bibinfo{person}{Michele Zorzi}.} \bibinfo{year}{2021}\natexlab{}.
\newblock \showarticletitle{Extending the ns-3 Spatial Channel Model for
  Vehicular Scenarios}. In \bibinfo{booktitle}{\emph{Proceedings of the 2021
  Workshop on ns-3}}. \bibinfo{publisher}{Association for Computing Machinery},
  \bibinfo{address}{Virtual Event, USA}.
\newblock


\bibitem[Zugno et~al\mbox{.}(2020)]%
        {tommaso:20}
\bibfield{author}{\bibinfo{person}{Tommaso Zugno}, \bibinfo{person}{Michele
  Polese}, \bibinfo{person}{Natale Patriciello}, \bibinfo{person}{Biljana
  Bojovi\'{c}}, \bibinfo{person}{Sandra Lag\'{e}n}, {and}
  \bibinfo{person}{Michele Zorzi}.} \bibinfo{year}{2020}\natexlab{}.
\newblock \showarticletitle{{Implementation of a Spatial Channel Model for
  ns-3}}. In \bibinfo{booktitle}{\emph{Proceedings of the 2020 Workshop on
  ns-3}}. \bibinfo{publisher}{Association for Computing Machinery},
  \bibinfo{address}{Gaithersburg, MD, USA}.
\newblock


\end{thebibliography}

\end{document}